\documentclass[aps,prx,twocolumn,showpacs,superscriptaddress,longbibliography]{revtex4-1}

\usepackage{amssymb}
\usepackage{amsfonts}
\usepackage{amsmath}
\usepackage{bm}
\usepackage{textcomp}
\usepackage{color}
\usepackage{longtable}
\usepackage{graphicx}
\usepackage{dcolumn}
\usepackage[a4paper=true,pagebackref=false]{hyperref}
\usepackage[normalem]{ulem}

\begin{document}

\title{Heat hunting in freezer: Direct measurement of quasiparticle diffusion in superconducting nanowire}

\author{M.~Zgirski}
\email{zgirski@ifpan.edu.pl}
\affiliation{Institute of Physics, Polish Academy of Sciences, Aleja Lotnikow 32/46, PL 02668 Warsaw, Poland}

\author{M.~Foltyn}
\affiliation{Institute of Physics, Polish Academy of Sciences, Aleja Lotnikow 32/46, PL 02668 Warsaw, Poland}

\author{A.~Savin} \affiliation{Low Temperature Laboratory, Department of Applied Physics, Aalto University School of Science, P.O. Box 13500, 00076 Aalto, Finland}

\author{A.~Naumov} \affiliation{Institute of Physics, Polish Academy of Sciences, Aleja Lotnikow 32/46, PL 02668 Warsaw, Poland}

\author{K.~Norowski} \affiliation{Institute of Physics, Polish Academy of Sciences, Aleja Lotnikow 32/46, PL 02668 Warsaw, Poland}

\date{\today}
\begin{abstract}
Propagation and relaxation of nonequilibrium quasiparticles in superconductors are of key importance for functioning of numerous nanoscale devices, enabling operation of some of them, and limiting the performance of the others. The quasiparticles heated above lattice temperature may relax locally via phonon or photon emission channels, or diffuse over appreciable distances in a nanostructure altering the functionality of their remote components. Tracing quasiparticles experimentally in real-time domain has remained the challenging task owing to their rapid dynamics. With electronic nanothermometry, based on probing of the temperature-dependent switching current of a superconducting nanobridge, we monitor heat pulse carried by a flux of nonequilibrium quasiparticles as it passes by our detector with a noise-equivalent temperature of 10$\,$mK/$\sqrt N$, where $N$ is the number of pulses probing the bridge (typically $N=10000$), and temporal resolution of a single nanosecond. The measurement provides the picture of quasiparticle diffusion in a superconducting aluminum strip and direct determination of the diffusion constant $D$ equal to 100\,cm$^2$/s with no energy dependence visible.
\end{abstract}

\maketitle

\begin{figure}[t]
\centering
\includegraphics[width=0.5\textwidth]{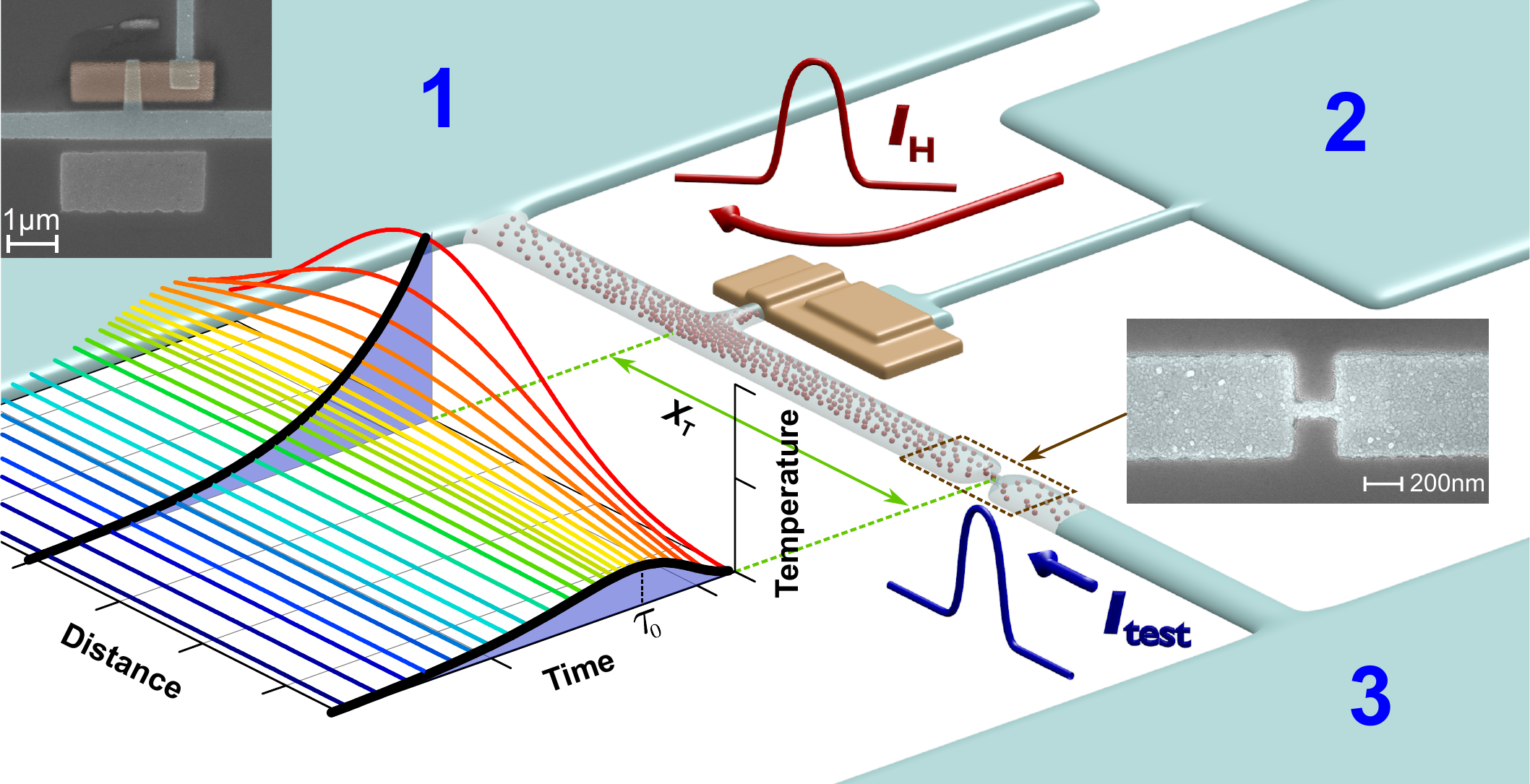}
\caption{\label{fig:Sample}Pictorial layout of the experiment. Hot electrons are created in the heater by applying short current pulse $I_H$ ($\sim$10\,ns long) flowing between ports 1 and 2. QPs start diffusing along the nanowire. Qualitatively, their population at the bridge location is derived from the time-evolving Gaussian profile as $N_T(\tau) \sim \frac{1}{\sqrt{2 \pi D \tau}} exp⁡(-\frac{X_T^2}{2D\tau})$ within the free-particle diffusion model, with the maximum of hot electron signal indicated at $\tau_0=X_T^2/D$. Similarly, hot electron population in the heater is $N_H(\tau) \sim \frac{1}{\sqrt{2 \pi D \tau}}$. Taking into account annihilation of nonequilibrium QP due to interaction with phonons and "colder" QPs allows to describe the temperature variations in the bridge quantitatively (see description in the text), but introduces only minor changes to temporal dynamics with almost the same time marking the onset and the maximum of the QP signal at the bridge. The insets show the SEM images of the copper heater and the aluminum nanobridge of the measured nanostucture. The testing pulse $I_{test}$, flowing between ports 1 and 3 is used to test the bridge temperature (see description in the text).}
\end{figure}

\begin{figure}[!]
\centering
\includegraphics[width=0.5\textwidth]{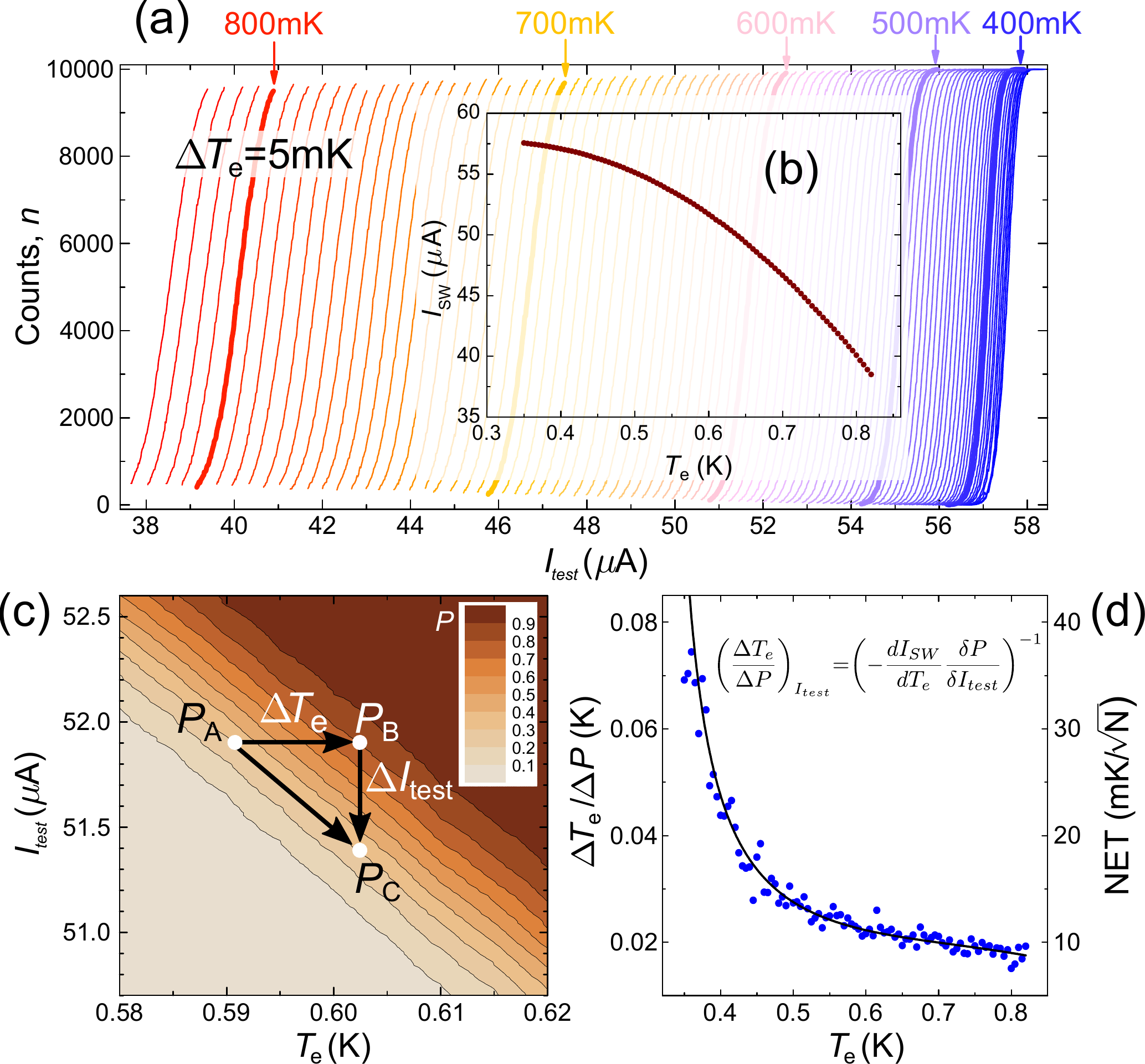}
\caption{\label{fig:Calibration}Switching thermometry. (a) Main panel presents collection of \textsf{S} curves recorded at various bath temperatures with $N=10000$. (b) Temperature dependence of the switching current $I_{SW}=I_{test}(P=0.5)$ as extracted from the \textsf{S} curves. The dependence serves as the calibration curve in the "Temperature from switching current" method. (c) Experimental dependence of the switching probability $P$ on the testing current $I_{test}$ and temperature $T_e$. The map is different presentation of \textsf{S} curves displayed in (a) over a limited range of temperatures. The zero change in probability on the path ABC ($\Delta P_{ABC}=\Delta P_{AB}+ \Delta P_{BC}=0$) allows to establish temperature responsivity $\Delta T_e / \Delta P$ as derived from the slope of \textsf{S} curves $\delta P / \delta I_{test}$ and switching current responsivity $dI_{SW} / dT_e$. (d) $\Delta T_e/\Delta P$ serving as the calibration curve in the "Temperature from probability" method (left axis) and the resulting NET (right axis). The lines imposed on the experimental data are empirical polynomial fits.}
\end{figure}

\begin{figure}[!]
\centering
\includegraphics[width=0.45\textwidth]{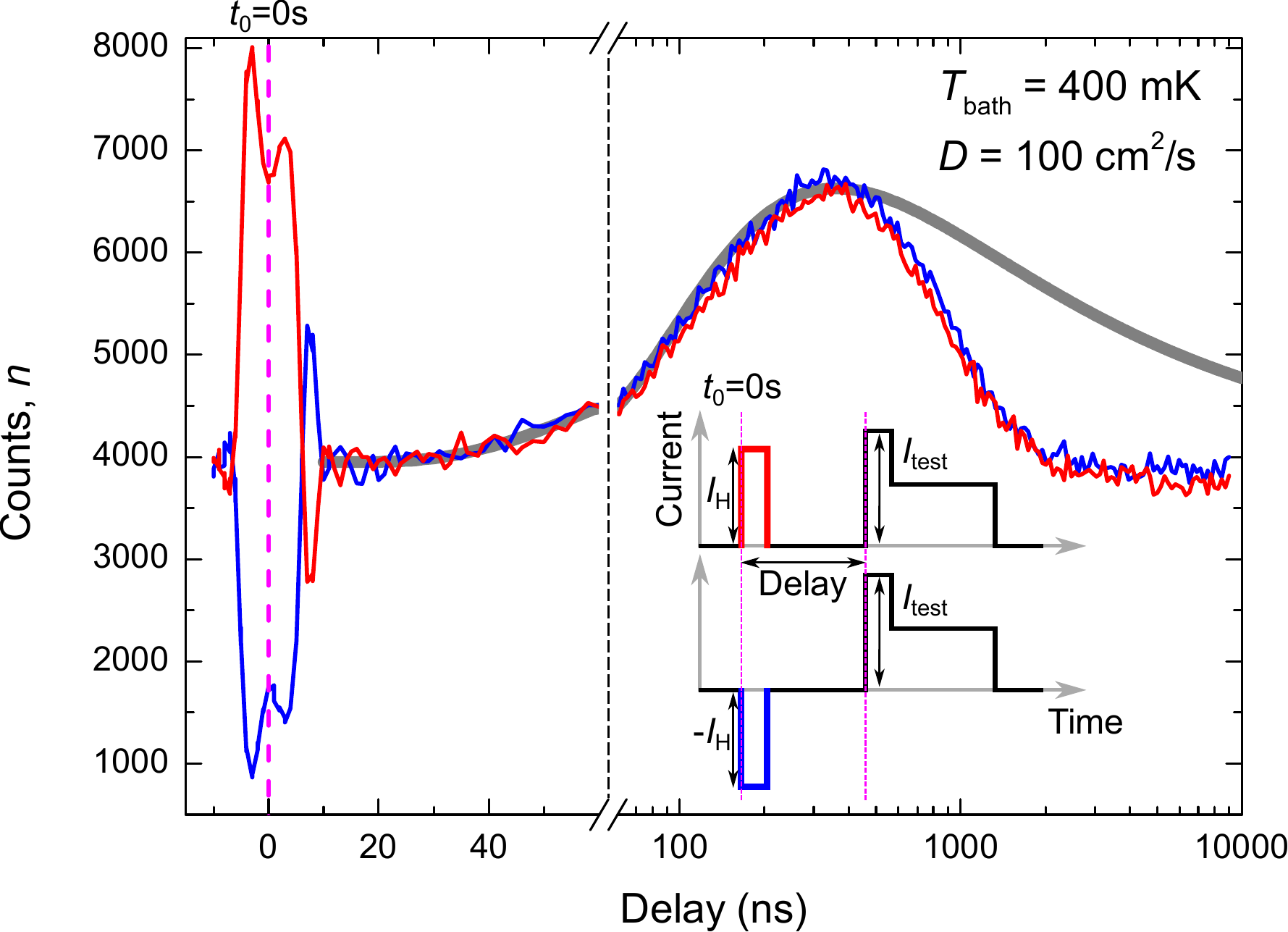}
\caption{\label{fig:Test}Switching probability as a function of delay between heating (red or blue, $I_H$ applied between ports 1 and 2) and testing pulse (black, $I_{test}$ applied between ports 1 and 3). Two traces are imposed on each other corresponding to two current polarities of the heating pulse. The delay equal 0 corresponds to the testing and heating pulse arriving to the device at the same time. In such case, small fraction of the heating pulse ($<0.5\%$) flows through the bridge, subtracting from or adding to the testing pulse. The temporal overlap of the heating and testing pulse is therefore well visible in the switching probability offering a convenient mean for timing calibration of the experiment. Noteworthy, two heating pulses of different polarities, but of the same amplitude, produce the same thermal response, as expected. The number of pulses to measure each point is $N=10000$ with repetition time of $100\,\mu$s. The thick solid line is the simple free-particle diffusion model presented in Fig.\,\ref{fig:Sample} fitted to the onset of the signal for $D=100$\,cm$^2$/s.}
\end{figure}

\begin{figure}
\centering
\includegraphics[width=0.45\textwidth]{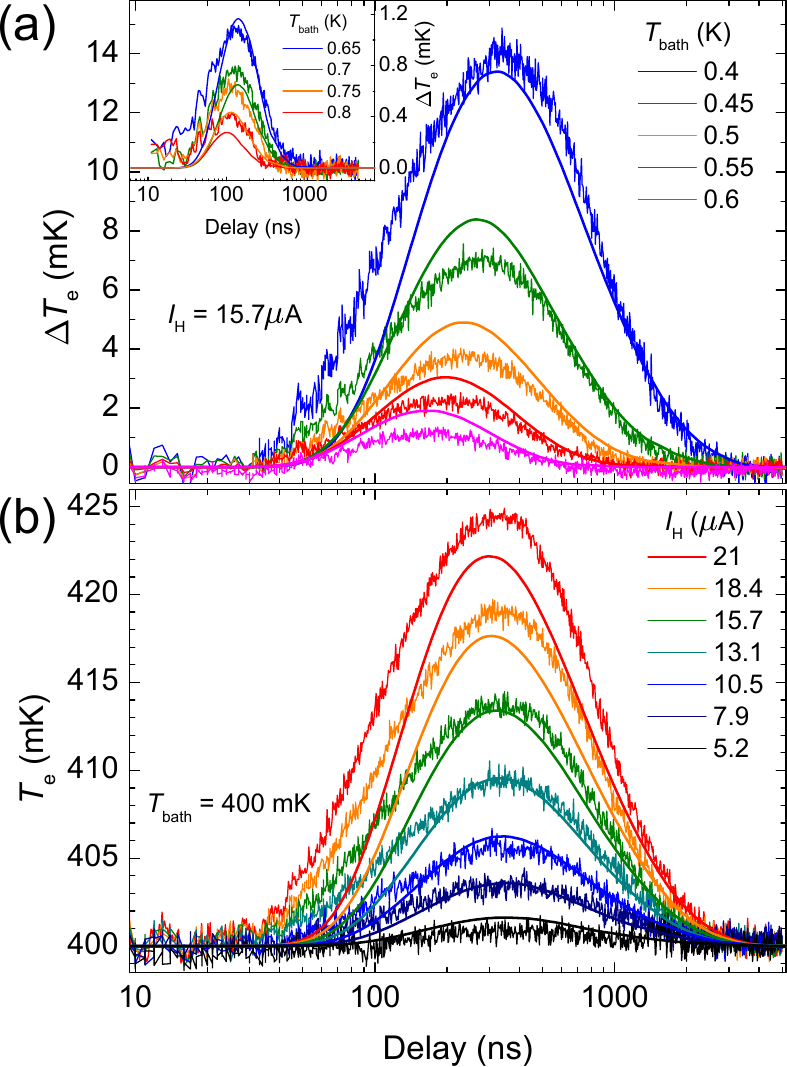}
\caption{\label{fig:Results}The temperature dynamics of the superconducting nanobridge after creating nonequilibrium QPs in the copper heater placed 60$\,\mu$m away with a short heating pulse. (a) Hot electron signal measured for the same heating pulse at various bath temperatures: 0.4-0.6$\,$K (main panel) and 0.65-0.8$\,$K (inset). Noteworthy, the hot-electron signal for $T=800\,$mK shows only $400\,\mu$K peak with accuracy better than $100\,\mu$K. (b) Hot electron signal measured at constant temperature for various heating pulses. The "noisy" profiles are experimental data for which the temperature is extracted with the "temperature from probability" procedure. Solid lines are calculated numerically for 1D heat flow model discussed in the text.}
\end{figure}

\emph{Introduction.}$-$The superconducting state at a finite temperature is characterized by equilibrium population of not paired electrons, known for their finite lifetime and tendency to recombine back into Cooper pairs as quasiparticles (QPs). When a metallic nanostructure is cooled down towards absolute zero, energy transfer between electrons and phonons becomes much less efficient resulting in thermal decoupling of the two systems\cite{Clarke1994}. If electrons absorb energy due to the Joule heating or irradiation with photons, they acquire temperature higher than that of the lattice. Such overheated electrons, often referred to as hot electrons, in a superconductor are known as nonequilibrium QPs. They diffuse in a nanostructure until they emit phonons\cite{Pekola2009} or photons\cite{Pekola2006} and adopt equilibrium occupation of states corresponding to the lattice temperature. The diffusion process, albeit much slower than ballistic propagation of electrons with the Fermi velocity $v_f$, has been too fast for existing experimental techniques to be traced in real-time domain. Dynamical thermal properties of nanostructures at low temperatures were mostly investigated by assumption of heat flow models describing thermal steady states, involving thermometry based on normal-metal-insulator-superconducting tunnel junctions\cite{Ullom1998, Pekola2008,Giazotto2012}, measurement of SQUID noise\cite{Jezouin2013} and Coulomb blockade in quantum dots\cite{Gasparinetti2011}. Since thermal and electrical attributes are intimately related, it was possible to get access to some thermal parameters by performing electrical transport measurements e.g. the Einstein formula for a degenerate conductor relates a diffusion constant and electrical conductivity\cite{Ullom1998}, the Wiedemann-Franz law relates electrical and thermal conductivities\cite{Betz2012,Michon2018}. There were also successful measurements of the thermal transients with temperature sensors embedded in RF or microwave resonators\cite{Cleland2004,Zgirski2016} with a noise equivalent temperature (NET) ranging from 90$\,\mu$K/$\sqrt {Hz}$\cite{Gasparinetti2015} to 10$\,\mu$K/$\sqrt {Hz}$\cite{Zgirski2016} respectively. They demonstrated real-time traces of the electron temperature for QPs releasing their excess energy to phonons. Owing to a typical bandwidth of 10$\,$MHz, experimentalists were able to trace thermal relaxation times down to $\sim\,$300ns at temperatures below ~0.5K\cite{Klapwijk2008,Pekola2018}. Utilizing recently developed switching thermometry with Josephson junction\cite{Zgirski2018}, we present direct measurement of the QP diffusion in the superconducting nanostructure achieving resolution below 100$\,\mu$K. Our study shows that tracing thermal processes in nanoscale with nanosecond resolution is possible and opens new perspectives for investigation of thermodynamics of low temperature quantum circuits. A proper understanding of thermal transients is essential for failure-free functioning of cryogenic nanodevices, involving design and development of nanoscale calorimeters and bolometers\cite{Wei2008,Govenius2016,Walsh2017}, microcoolers\cite{Savin2006} and qubits. Devices like single electron boxes, proposed as building blocks of modern current standard, suffer from the presence of QPs responsible for leakage currents and resulting $"$counting errors$"$, that spoil metrological applications. Similarly, microcoolers' performance is degraded due to the QP poisoning. The QPs are also known to have a detrimental influence on the coherence times of superconducting\cite{Serniak2018} or Majorana\cite{Rainis2012,Aasen2016} qubits. On the other hand, the creation of QPs due to photon absorption makes it possible for superconducting bolometers to detect incident radiation with the lifetime of QPs imposing an intrinsic limitation on the bandwidth of such sensors\cite{Klapwijk2008}. Our study may offer new ways for advancement of the emergent fields of quantum thermodynamics\cite{Pekola2015} and phase-coherent caloritronics\cite{Fornieri2017}. The latter involves generation and manipulation of heat currents to demonstrate novel-concept devices\cite{Giazotto2017,Giazotto2014}. Harnessing heat current pulses as thermal counterparts of electrical signals could extend the discipline beyond steady-states investigations and provide a competitive alternative for phononics\cite{Li2012} and spin caloritronics\cite{Bauer2012}.

\emph{Sample.}$-$We have fabricated a device with a normal metal heater galvanically connected to a superconducting aluminum nanowire interrupted with a nanobridge (Fig.\,\ref{fig:Sample}). The bridge is a sensitive thermometer and the Joule heated copper island, placed $X_T=60\mu$m away, is a source of nonequilibrium QPs, where electrons are promoted to higher energy levels with the local Fermi-Dirac distribution characterized by temperature elevated above the phonon temperature. The distribution relaxes towards equilibrium with phonons only gradually and relaxation process may require seconds in a few milikelvin temperature\cite{Roukes1985,Gershenson2001,Huard2007}. Hot electrons move with $v_f$ (equal to a few percent of speed of light) but due to scattering on different lattice defects, i.e. grain boundaries, sample surface or impurities, their spreading in nanostructure is not so fast but instead takes on diffusive character. Qualitatively, in its random walk an average hot electron bounces off each $\sim 2-100\,$nm (a length known as an elastic mean free path $l_{mfp}$) and after many collisions covers distance given by Einstein-Smoluchowski law: $\left\langle l^2 \right\rangle=D\tau$, where $\left\langle l^2 \right\rangle$ is a mean square displacement from a starting point after time $\tau$ and $D$ is the diffusion constant. It accounts for $\sim 1-1000\,$ns required for hot electrons to spread in a conventional microstructure with size $10-100\,\mu$m. For presented sample hot electrons diffuse along the wire transporting the heat away from the copper island. On the way they lose energy to phonons ($\dot{q}_{ep}$ is the energy flux to phonons) and equilibrate with local QPs occupying lower energy states characterized by lower temperature. We assume that in each section of the wire electrons are described with equilibrium Fermi-Dirac distribution and their temperature is well-defined. The energy required for the hot electron flux to equilibrate with local electrons is accounted for by the heat capacity $C_p$ of the QPs in the superconducting state. Finally, QPs arrive to the nanobridge, whose switching current is sensitive to their local population.

\emph{Switching thermometry.}$-$We use and further develop nanosecond thermometry based on stochastic switching of a Josephson junction from superconducting to normal state\cite{Zgirski2018, Zgirski2019}. A particular type of Josephson junction, a superconducting aluminum nanobridge known in literature as the Dayem bridge, is well tailored for tracing rapid changes in temperature, which are expected when hot electrons propagate across the nanostructure. The bridge is probed with train of $N$ current pulses (see $I_{test}$ pulse send between port 1 and 3 in Fig.\,\ref{fig:Sample}). In response to each pulse it may either remain in the superconducting state or transit to the normal state. The switching process is both current and temperature dependent. Number of switching events $n$ increases with amplitude of the probing pulse and switching probability $P=n/N$ renders familiar S-shaped curve. Such \textsf{S} curve is centered at lower current amplitudes for higher temperatures (Fig.\,\ref{fig:Calibration}(a)). The variation of \textsf{S} curve position with temperature (Fig.\,\ref{fig:Calibration}(b)) and its slope define the temperature responsivity $\Delta T_e/\Delta P$ at constant testing current amplitude (Fig.\,\ref{fig:Calibration}(c,d)). Alternatively, temperature can be derived by associating unique current amplitude corresponding to $P=0.5$ switching probability with temperature (Fig.\,\ref{fig:Calibration}(b)). We call this two methods of the switching thermometry "Temperature from probability" and "Temperature from switching current", respectively. Metrological aspects of both of them are described in Supplemental Material\cite{Supplemental_Hot_electrons}.

\emph{QP diffusion measurement.}$-$To trace propagation of hot electrons, we first create their population applying short current pulse ($\leq$10$\,$ns) to the copper island (port 1 and 2 in Fig.\,\ref{fig:Sample}) and then, after a few dozens of nanoseconds, we send the testing pulse on the bridge (port 1 and 3 in Fig.\,\ref{fig:Sample}). We repeat the whole sequence $N$ times to measure switching probability corresponding to the given delay between the two pulses. The delay can be set with accuracy better than 1$\,$ns. Varying the delay allows to reconstruct the temporal variation of the switching probability as electron diffusion proceeds. The temporal resolution of the measurement is limited by the length of the probing portion of the testing pulse (only topmost part of the Gaussian-shaped pulse with $FWHM\cong8\,$ns can make the bridge switch) and approaches a single nanosecond in our experiment. The typical experimental profile is presented in Fig.\,\ref{fig:Test}. The hot-electron signal peaks up $\sim 300\,$ns after application of the 10$\,$ns long heating pulse which qualitatively agrees with diffusion time across 60$\,\mu$m long nanowire, discussed earlier. One can observe the delay of $\sim 40\,$ns between the heating pulse and the onset of the signal. Importantly, the delay shows that switching current of the bridge depends on the local distribution (local temperature) of QPs. The experimental profile allows us to determine the diffusion constant via direct comparison with simple free-particle diffusion model (Fig.\,\ref{fig:Test}, see also Supplemental Material\cite{Supplemental_Hot_electrons}). The fit yields value of $D=(100 \pm 5)\,$cm$^2$/s. Since the diffusion constant, governing the spreading of electrons in a one particular direction (i.e. along the length of the wire), is equal to $D=1/3 \cdot v_f \cdot l_{mfp}$, setting $v_f=2 \cdot 10^6\,$m/s we obtain $l_{mfp}=15$\,nm, a value comparable with the grain size of our polycrystalline aluminum (see SEM photo in Fig.\,\ref{fig:Sample}). Using calibration dependence $\Delta T_e/\Delta P$ (Fig.\,\ref{fig:Calibration}(d)), we convert the measured signal into electron temperature. In Fig.\,\ref{fig:Results} we present results of such conversion for different bath temperatures and heating currents. We have also heated the copper island with pulses of different duration observing gradual build-up of the temperature profile until the steady-state with an elevated temperature on the bridge was reached\cite{Supplemental_Hot_electrons}. Similarly, we have collected transients appearing after 10$\,\mu$s-long heating pulse is turned off\cite{Supplemental_Hot_electrons}.

\emph{Thermal modeling.}$-$The onset of the QPs is well described by the free-particle diffusion model which nevertheless fails to explain the observed signal at longer delays. To understand the overall shape of experimental diffusion profile we elaborate a more detailed thermal model describing evolution of temperature in the wire. Firstly, owing to enhanced electron-phonon coupling, hot QPs are expected to dissipate their energy to phonons before they reach the bridge. Secondly, diffusing electrons should lose some energy to equilibrate with local and "colder" QPs. We map our three-terminal device into 1D model to perform simplified heat flow analysis. Instead of considering leads 1 and 2 we replace them with a single lead of the same cross-section as lead 3.

We analyze the diffusion process by numerically solving one-dimensional time dependent heat flow equation:
\begin{equation*}
\cfrac{d}{dx}\left(\kappa(T_e)\cfrac{dT_e}{dx}\right)=C_p(T_e)\cdot \cfrac{\delta T_e}{\delta t}+\dot{q}_{ep}(T_e)-\cfrac{r(x)\cdot I_H(t)^2}{S}
\end{equation*}
where left part of equation describes the net heat flux carried by hot electrons ($\kappa(T_e)$ is the electron thermal conductivity) and terms on the right hand side describe increase of electron energy ($C_p(T_e)$ is electron heat capacity), power transmitted to phonons $\dot{q}_{ep}$ and heating confined spatially to the heater stripe of 3$\,\mu$m length. $I_H(t)$ defines a time-dependent current pulse, $r(x)$ is the resistance per unit length and $S=600\,$nm $\times$ 30$\,$nm is the cross-section of the aluminum nanowire.

In solving the equation we assume the literature-based values of thermal parameters for aluminum nanowire on one side of the heater (where thermometer is placed) and the values rescaled by factor $k=3$ for the second side\cite{Supplemental_Hot_electrons}. We find the best correspondence between numerical simulation and experimental data assuming the effective resistance of the copper heating island $R_{eff}=1.6\,\Omega$. This parameter is used consequently for all modeling. The $R_{eff}$ value is roughly three times smaller than the measured resistance ($R=4.5\,\Omega$) of the heater line spanning between ports 1 and 2 (see Fig.\,\ref{fig:Sample}). The difference can be ascribed to the fact that hot electrons created in the copper heater are Andreev-reflected at the normal metal-superconductor interface and only those with sufficiently high energies can enter into aluminum nanowire as nonequilibrium QPs\cite{Vinokur2003, Chandrasekhar1998}. Also, a reasonable modification of material parameters would result in the higher fitted value of $R_{eff}$, reducing, to some extent, its departure from $R$. The calculated temperature profiles are imposed on the experimental data in Fig.\,\ref{fig:Results} and in Supplemental Material\cite{Supplemental_Hot_electrons}.

The measurements presented in Fig.\,\ref{fig:Results} reveal spatial range of QPs, their lifetime and identifies mechanisms responsible for their annihilation. We observe the fast build-up (of order of 100$\,$ns) of QPs population at the detector in response to a remote heating pulse. The signal is much more pronounced at lower temperatures, where electron-phonon coupling is suppressed\cite{Supplemental_Hot_electrons}. At $T_{bath}=0.4\,$K relaxation time is of order of 1$\,\mu$s and it decreases to $\sim 100\,$ns at $T_{bath}=0.8\,$K owing to dominant role of electron-phonon coupling on electron temperature relaxation at higher temperatures.

\emph{Discussion.}$-$ One could expect that at higher bath temperatures, owing to higher average energy QPs should arrive to the detector faster. Such expectation is a result of group velocity $v_g$ scaling with energy as $v_g=v_f \sqrt{1-\Delta^2/E^2}$, where $\Delta$ is the superconducting gap\cite {Ullom1998}. It is not what we see in the experiment. The temporal onset of the signal does not depend on temperature and heating power. Instead, all profiles for short delays can be fitted with the same diffusion constant \cite{Supplemental_Hot_electrons}. The long lifetime and spatial range of QPs at low temperatures make it obligatory to engineer gap and trap structures for single electron boxes\cite{Aumentado2004,Taupin2016}, microcoolers\cite{Manninen2000} and qubits\cite{Martinis2009}. The presented experiment could be easily modified to test the efficacy of QP trapping in normal metal\cite{Ullom2000,Riwar2016,Courtois2012} or in Meissner or vortex states\cite{Taupin2016}, if normal metal island or a wider piece of aluminum strip (allowing to accommodate a vortex\cite{Martinis2004}) was inserted on the way between the heater and the detector. We measured QPs propagation down to 400$\,$mK. The lifetime and propagation range are expected to be vastly increased when lowering temperature towards absolute zero. The natural extension of the current work is a measurement of QP diffusion at temperatures below 100$\,$mK, typical for superconducting qubit operation. It can be accomplished by using Josephson junction exhibiting switching current sensitivity at lower temperatures compared to the presented aluminum nanobridge. An SNS proximity junction or a titanium nanobridge would be perhaps good candidates. Studies at lowest temperatures could help to resolve the mystery of the residual QP density, which appears not to follow the BCS theory.

\emph{Conclusions.}$-$We demonstrate the real-time measurement of the nonequilibrium QP diffusion in the superconducting aluminum nanowire. Such investigation is possible because our fast thermometry delivers resolution at single nanosecond level ($t_{res} \sim 1\,ns$) accessing for the first time the regime where $t_{res} \ll L^2/D$ with $L$ being the spatial extent of the experiment (i.e. distance between QP source and detector). Our data are in agreement both with the simple model of the free-particle diffusion (allowing for direct determination of the diffusion constant), and more involved thermal model taking into consideration the electron-electron and electron-phonon scatterings with the first mechanism being accounted for by the electron heat capacity term and the second one by electron-phonon coupling in the heat flow equation. Curiously enough, the method involves measurement of somewhat abstract probability from which electron temperature can be obtained. The presented switching thermometry allows to study fast nonequilibrium thermal processes in nanostructures offering an attractive tool for experimental quantum thermodynamics and caloritronics.

The authors thank Olli-Pentti Saira and Denis Vodolazov for helpful discussions and Paulina Grzaczkowska for a technical support. The work is financed by Foundation for Polish Science (First TEAM/2016-1/10).

\newpage

\section*{"Heat hunting in freezer: Direct measurement of quasiparticle diffusion in superconducting nanowire" Supplemental Material}



\begin{figure*}[!]
\centering
\includegraphics[width=0.85\textwidth]{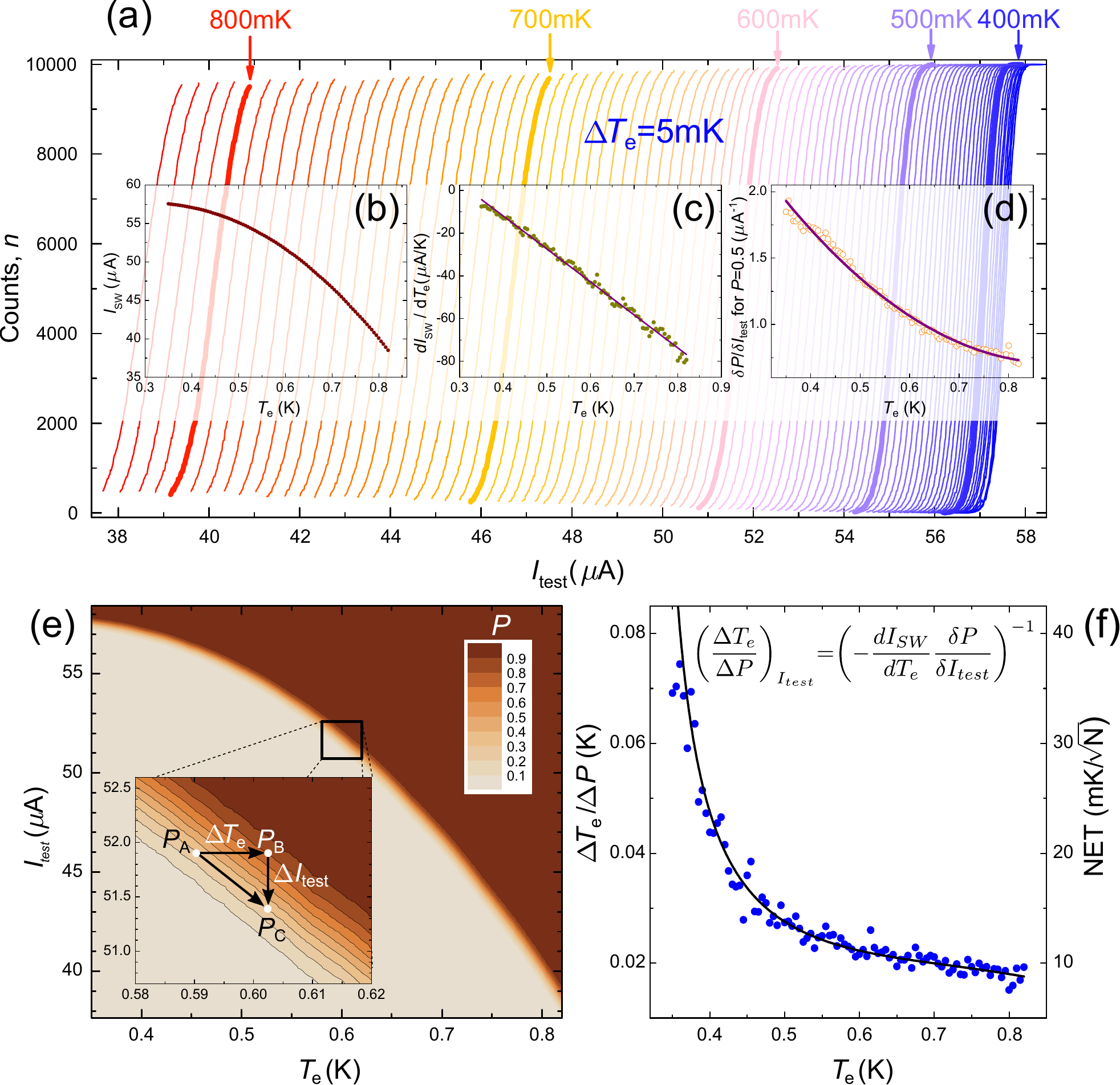}
\caption{\label{fig:Calibration_full}Switching thermometry. (a) Main panel presents collection of \textsf{S} curves recorded at various bath temperatures with $N=10000$. (b) Temperature dependence of the switching current $I_{SW}=I_{test}(P=0.5)$ as extracted from the \textsf{S} curves. The dependence serves as the calibration curve in the "Temperature from switching current" method (see discussion in the text). (c) Switching current responsivity $dI_{SW}/dT_e$. (d) Slope of the \textsf{S} curves $\delta P/\delta I_{test}$. (e) Experimental dependence of the switching probability $P$ on the testing current $I_{test}$ and temperature $T_e$. The map is different presentation of \textsf{S} curves displayed in (a). Inset: Close-up of the dependence with indicated probabilities $P_A$, $P_B$ and $P_C=P_A$. The zero change in probability on the path ABC ($\Delta P_{ABC}=\Delta P_{AB}+ \Delta P_{BC}=0$) allows to establish temperature responsivity $\Delta T_e / \Delta P$. (f) $\Delta T_e/\Delta P$ serving as the calibration curve in the "Temperature from probability" method (left axis) and the resulting NET (right axis). The lines imposed on the experimental data are empirical polynomial fits.}
\end{figure*}

\begin{figure*}[!]
\centering
\includegraphics[width=0.85\textwidth]{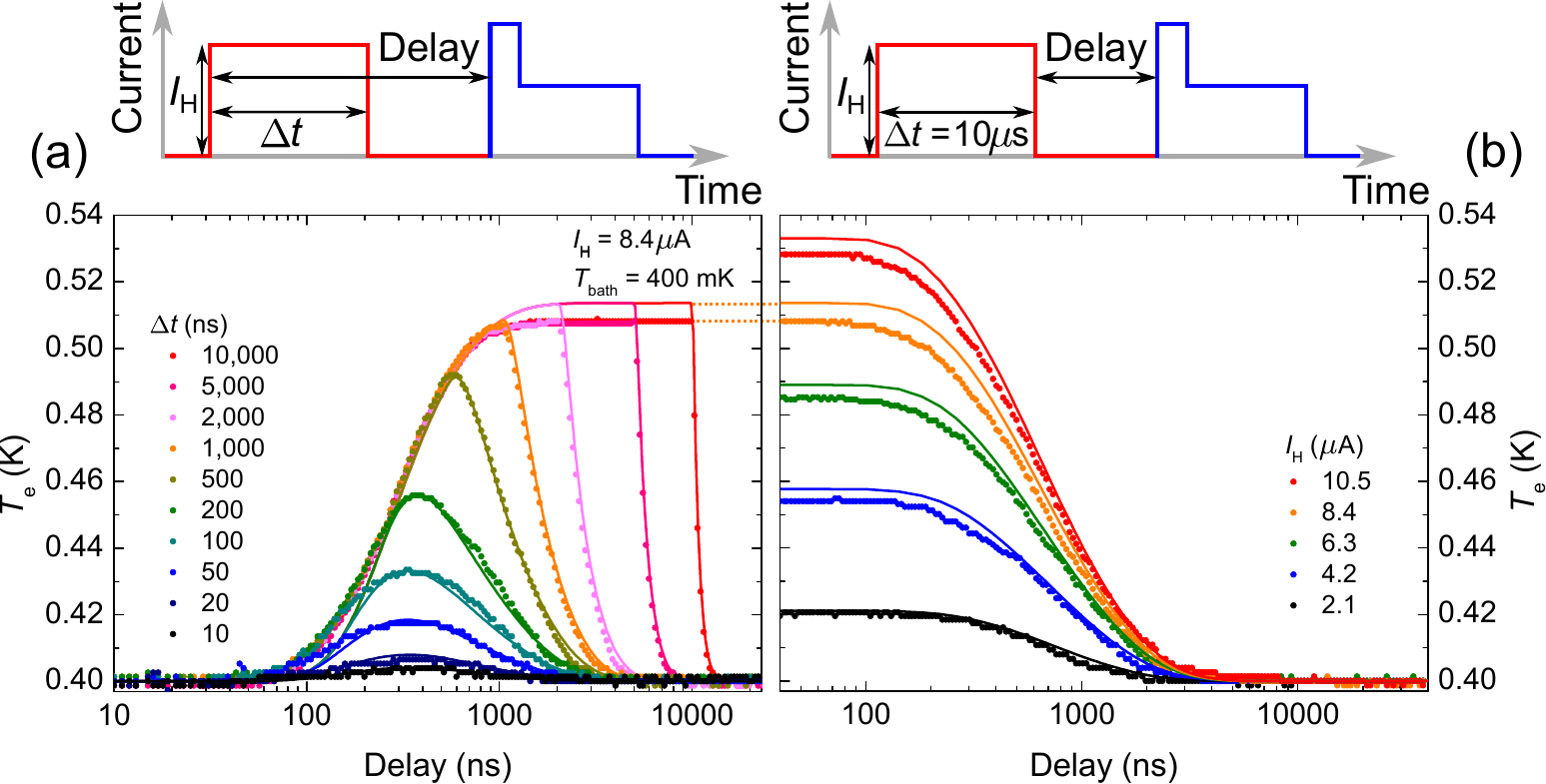}
\caption{\label{fig:Results_SS}Temperature dynamics of the superconducting nanobridge after creating nonequilibrium QPs in Cu heater placed 60$\,\mu$m away with a long heating pulse. (a) Hot-electron signal measured for heating pulses of the various duration and the same amplitude ($I_H=8.4\,\mu A$). (b) Hot-electron signal measured for heating pulses of the various amplitude and the same duration ($\Delta t = 10\,\mu s$). The definition of measurement is depicted above each figure. The "dotted" profiles are experimental data for which the temperature is extracted with the "temperature from switching current" procedure. Solid lines are calculated numerically for 1D heat flow model discussed in the text.}
\end{figure*}

\begin{figure*}[!]
\centering
\includegraphics[width=0.7\textwidth]{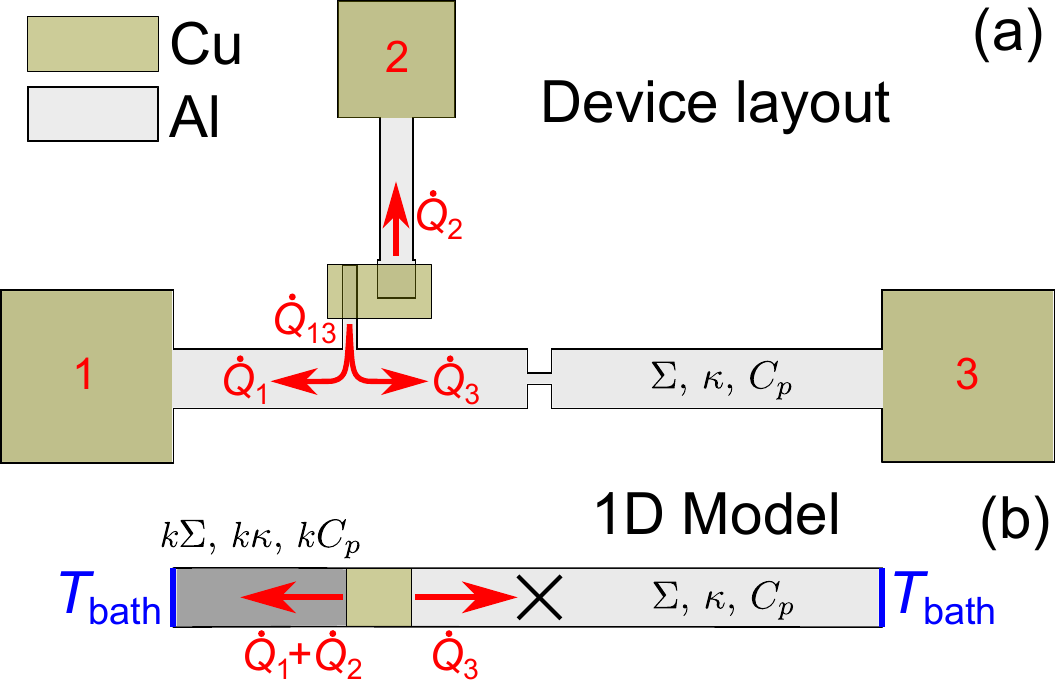}
\caption{\label{fig:Model}(a) Layout of the nanostructure. (b) The modeled 1D strip with left hand side thermal parameters modified to account for the heat flow asymmetry in the real structure. Basing on similar surfaces of NS interfaces defined on the top of the copper island (see inset of Fig. 1 in the main text) we assume that total generated heat gives rise to two equal heat fluxes $\dot{Q}_2$ and $\dot{Q}_{13}$ ($\dot{Q}_2=\dot{Q}_{13}=\dot{Q}_1+\dot{Q}_3$ and $\dot{Q}_1=\dot{Q}_3$) responsible for evacuation of the energy from the heater.}
\end{figure*}

\begin{figure*}[!]
\centering
\includegraphics[width=0.5\textwidth]{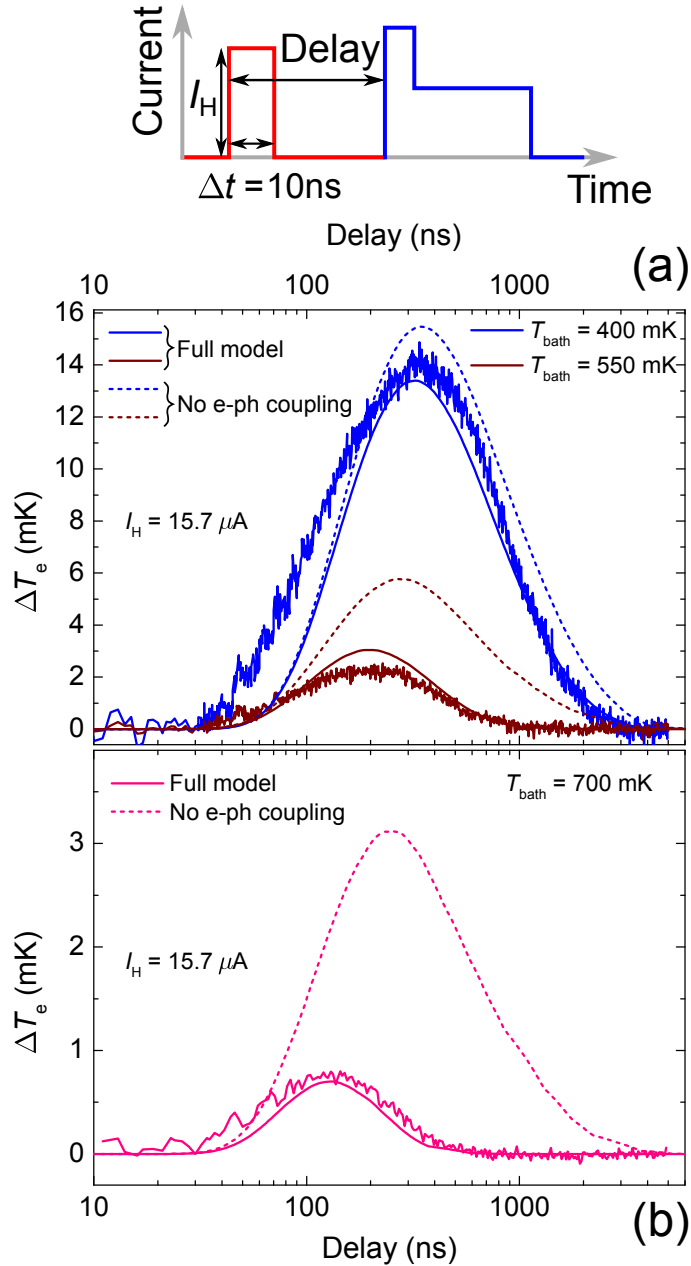}
\caption{\label{fig:studies1} (a) Hot electron signals measured for $T_{bath}=400\,$mK and $T_{bath}=550\,$mK with the sequence of pulses depicted above. Continuous lines are calculated numerically for model including QP diffusion and electron-phonon coupling. Dashed lines show model results for QP diffusion mechanism only. (b) Analogous results for $T_{bath}=700\,$mK.}
\end{figure*}

\section*{1. Sample fabrication}

The nanostructure (Fig.\,\ref{fig:Sample} in the main text) was fabricated with conventional electron-beam lithography followed by sequential deposition of 30$\,$nm of aluminum (at an angle of 0$\,$deg) and 30$\,$nm of copper (at an angle of 50$\,$deg) at a base pressure of $10^{-8}\,$mBar in the electron-beam evaporator. Deposited aluminum formed a long nanowire (L = 180$\,\mu$m) interrupted with the Dayem nanobridge in the middle, and a lead connecting port 2 with the copper island, placed 60$\,\mu$m away from the bridge. All three contact pads (ports 1,2 and 3) were covered with copper ensuring proper thermalization of electrons, owing to enhanced strength of electron-phonon coupling in copper compared to that of superconducting aluminum.

\section*{2. Experimental apparatus}
The switching measurements were performed in the bottom-loaded Triton 400 dilution refrigerator. We launched fast heating and probing pulses from two-port 80$\,$MHz Arbitrary Waveform Generator (Slave generator) triggered by 80$\,$MHz AWG 33250A (Master generator). The pulses were guided to the sample by means of HF lines with total attenuation of 56$\,$dB (testing line) and 53$\,$dB (heating line) distributed at 293$\,$K, 60$\,$K, 4$\,$K and mixing plate. The last sections of the HF lines were interrupted with 100$\,\Omega$ resistors used for measuring the electrical current. One filtered twisted pair was used to measure voltage drop across the 100$\,\Omega$ resistor to enable calibration of the current flowing through the junction (port 1 and 3). The another twisted pairs was connected between the input of the junction (port 1) and the sample holder ground to monitor voltage drop across the junction, and thus switching events. The similar arrangement of the two twisted pairs was used to control current and voltage of the heater. Twisted pairs were connected to room temperature amplifiers (NF LI-75A and DL Instruments 1201) to obtain total amplification of $\sim 2000$ with their outputs connected to LeCroy HRO 66Zi oscilloscope. The switching events were visible on the scope as voltage pulses exceeding a certain threshold and counted with the build-in function of the instrument. The whole experiment was controlled through LabView program responsible for triggering of the pulses, their timing, duration, amplitude, mutual shift etc., and communicating with the scope and temperature controller of the Triton.

\section*{3. Switching thermometry - metrology}

We use and further develop the recently introduced nanosecond thermometry based on stochastic switching of a Josephson junction from superconducting to normal state\cite{Zgirski2018_S, Zgirski2019_S}. A particular type of Josephson junction, a superconducting aluminum nanobridge known in literature as the Dayem bridge, is well tailored for tracing rapid changes in temperature, which are expected when hot electrons propagate across the nanostructure. The bridge is probed with train of $N$ current pulses. In response to each pulse it may either remain in the superconducting state or transit to the normal state. The switching process is both current and temperature dependent. Number of switching events $n$ increases with amplitude of the probing pulse and switching probability $P=n/N$ renders familiar S-shaped curve. Such \textsf{S} curve is centered at lower current amplitudes for higher temperatures (Fig.\,\ref{fig:Calibration_full}a). The variation of \textsf{S} curve position with temperature (Fig.\,\ref{fig:Calibration_full}b,c) and its steepness (Fig.\,\ref{fig:Calibration_full}d) define the temperature responsivity $\Delta T_e/\Delta P$ at constant testing current amplitude (Fig.\,\ref{fig:Calibration_full}f). Alternatively, temperature can be derived by associating unique current amplitude corresponding to $P=0.5$ switching probability with temperature (Fig.\,\ref{fig:Calibration_full}b). Below we describe these two methods in detail.

\subsection*{3a. Temperature from probability}
During a thermal transient, when a nanobridge is probed with pulses of a constant amplitude, excess values of the switching probability $\Delta P$ correspond to departures of electron temperature $\Delta T_e$ from lattice temperature $T_{ph}$. In the linear range of an \textsf{S} curve ($\delta P/\delta I_{test}=const$), covering the interval $0.2<P<0.8$, $\Delta P$ is proportional to $\Delta T_e$. We tune switching current to obtain $P=0.2$ for $T_e=T_{ph}$. Then electron temperature during any moment of relaxation is $T_e=T_{ph}+\Delta T_e=T_{ph}+(-\delta P/\delta I_{test} \cdot dI_{SW}/dT_e)^{-1}\Delta P$. The conversion formula is the result of mutual relation (i.e. triple product rule) between three partial derivatives $\delta P/\delta I_{test}$, $\delta I_{test}/\delta T_e$, $\delta P/\delta T_e$ illustrated in Fig.\,\ref{fig:Calibration_full}e. Collection of $P(I_{test})$ dependences at different temperatures gives knowledge of $\delta P/\delta I_{test}$ and $dI_{SW}/dT_e$, and allows to calculate the temperature responsivity at constant testing current amplitude $\left( {\Delta T_e}/{\Delta P} \right) _{I_{test}}$. The uncertainty in $T_e$ determination is set by accuracy of probability measurement\cite{Zgirski2019_S}, that is $\Delta P_{un}=[P(1-P)/N]^{1/2}$ (N - number of pulses) and reads $\Delta T_{e,un}=(-\delta P/\delta I_{test} \cdot dI_{SW}/dT_e)^{-1}\Delta P_{un}$. The method is applicable only for linear regime when $\Delta P \sim \Delta T_e$. We define a NET in units of K/$\sqrt{N}$ as the normalized uncertainty $\Delta T_{e,un} \cdot \sqrt N$ (see Fig.\,\ref{fig:Calibration_full}f, right axis). Increasing number of testing pulses is equivalent to reduction of the measurement bandwidth.

\subsection*{3b. Temperature from switching current}
Second way of temperature determination requires only knowledge of $I_{SW}(T_e, P=P_{goal})$ curve. Here, during relaxation process, the bisection algorithm iteratively finds the switching current corresponding to probability $P=P_{goal} \pm \Delta P_{bis}$ (the search is stopped when measurement yields the probability from the specified interval). The obtained value is converted into temperature. The uncertainty in $T_e$ determination is set by accuracy of probability measurement, like in the first method, but in addition it also suffers from non-zero value of $\Delta P_{bis}$, which for typical experiment is $\Delta P_{bis}=0.01-0.02$, significantly bigger than $\Delta P_{un}$. We get for the second method $\Delta T_{e,un}=(-\delta P/\delta I_{test} \cdot dI_{SW}/dT_e)^{-1} \cdot (\Delta P_{un}+\Delta P_{bis})$. The method is less sensitive but it is well-suited for probing transients in a nonlinear regime.

\section*{4. Long heating pulse response}

We perform studies for heating pulses with duration ranging from 10\,ns up to 10 $\mu$s to observe the emergence of the thermal steady state (Fig.\,\ref{fig:Results_SS}a). Subsequently, starting with the steady states for different heating powers we measure corresponding thermal relaxation curves (Fig.\,\ref{fig:Results_SS}b). In all cases we are able to reproduce experimental results with 1D heat flow model, discussed in the main manuscript, with the same material and geometrical parameters.

\section*{5. Modeling of the heat flow in the studied device}

We map our three-terminal device into 1D model to perform simplified heat flow analysis. Instead of considering leads 1 and 2 we replace them with a single lead of the same cross-section as lead 3, but with thermal parameters rescaled by a single geometry factor $k=\frac{\dot{Q}_1+\dot{Q}_2}{\dot{Q}_3}=3$ (Fig.\,\ref{fig:Model}).

\section*{6. Influence of the electron-phonon interaction on the quasiparticle diffusion in the aluminum nanowire}

\begin{figure*}[!]
\centering
\includegraphics[width=0.5\textwidth]{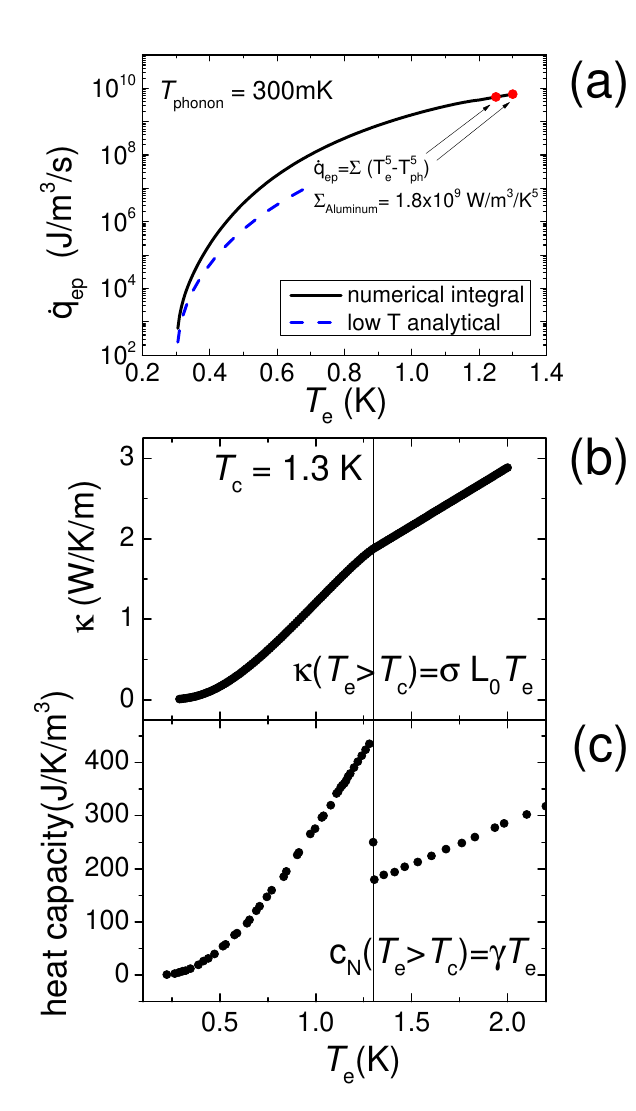}
\caption{\label{fig:parameters} Numerical parameters used to solve heat flow equation. (a) Electron-phonon coupling, (b) Thermal conductivity, (c) Heat capacity.}
\end{figure*}

At low temperatures electron-phonon interactions are weak and they become stronger with increasing temperature (see Fig.\,\ref{fig:parameters}a). It is expected that for lower temperatures quasiparticle (QP) diffusion is the dominant mechanism of heat transfer, while electron-phonon coupling becomes more important at higher temperatures. To compare the contribution of these two mechanisms we disabled electron-phonon coupling in the simulation leaving diffusion of QPs as the only relaxation channel. Temperature differences calculated with full (continuous lines in Fig.\,\ref{fig:studies1}) and reduced (dashed lines in Fig.\,\ref{fig:studies1}) model increase when bath temperature is higher. While at $T_{bath}=400\,$mK electron-phonon interaction is only a minor correction to overall heat flow (Fig.\,\ref{fig:studies1}a), at $T_{bath}=700\,$mK it attenuates QP signal significantly (Fig.\,\ref{fig:studies1}b).

\begin{figure*}[!]
\centering
\includegraphics[width=0.5\textwidth]{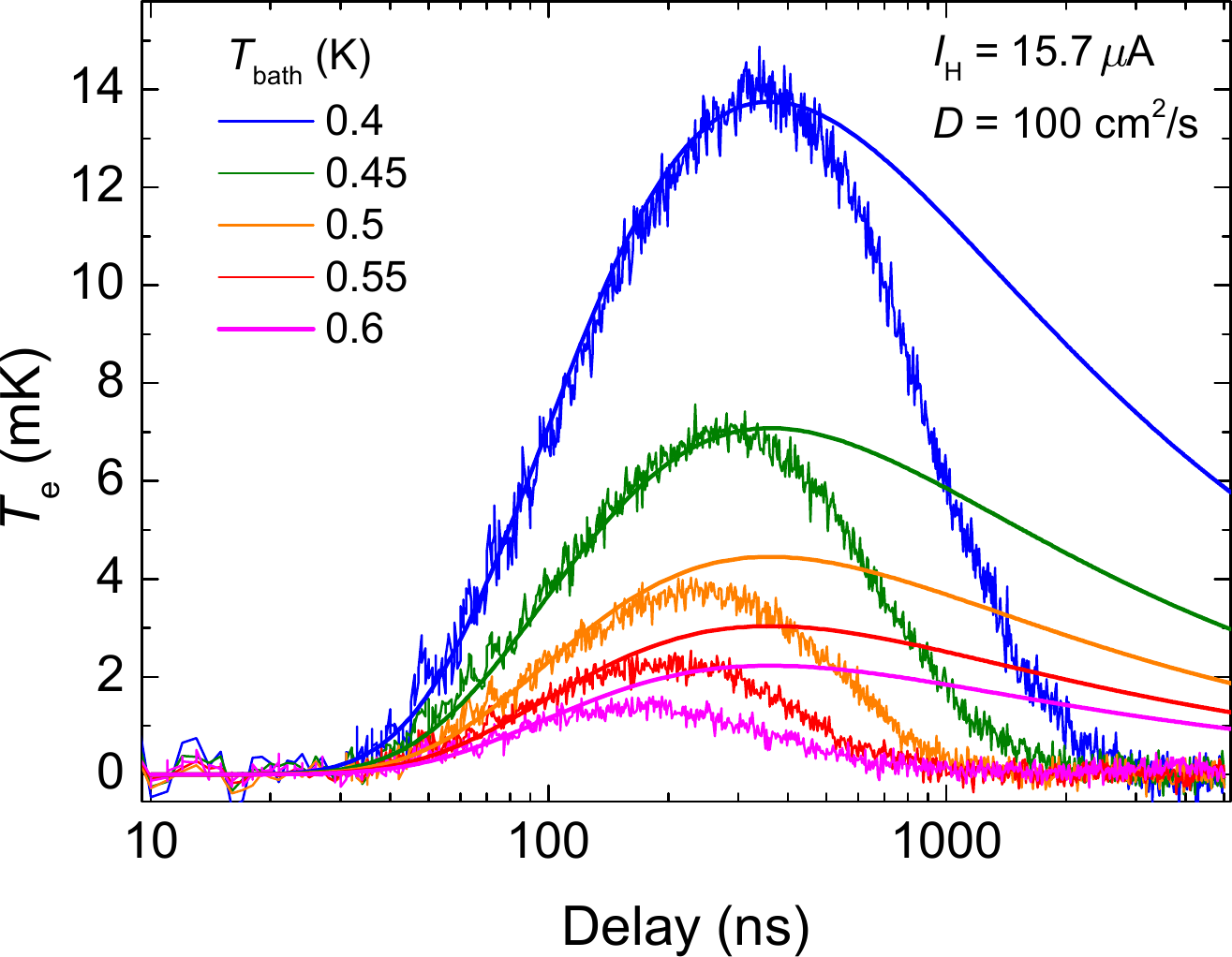}
\caption{\label{fig:Diffusion}Experimental profiles for different bath temperatures (the same as in the main text) and free-particle diffusion model fitted to the onset of the signals (solid lines).}
\end{figure*}

\begin{figure*}[!]
\centering
\includegraphics[width=0.5\textwidth]{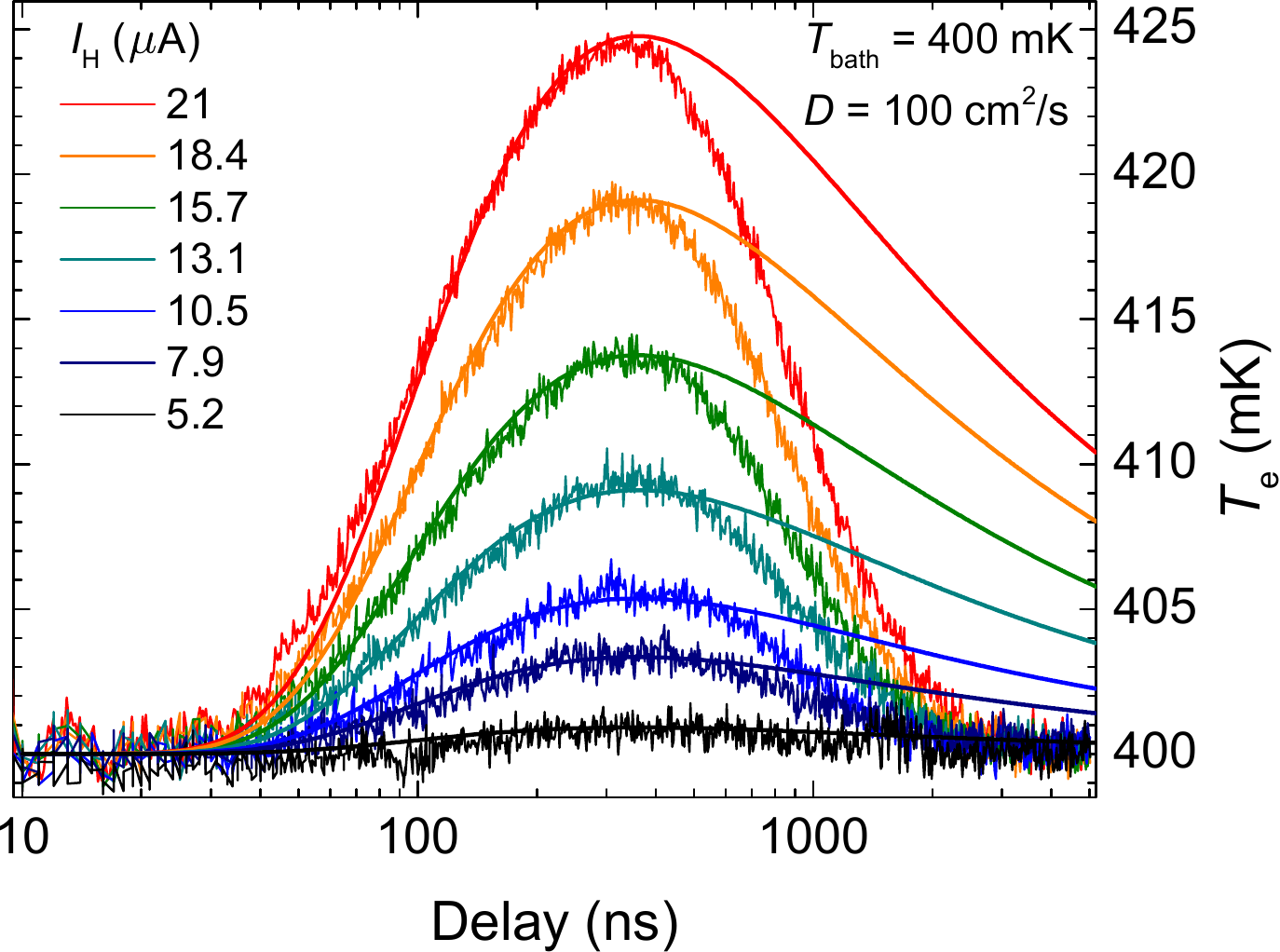}
\caption{\label{fig:Diffusion2}Experimental profiles for different heating currents (the same as in the main text) and free-particle diffusion model fitted to the onset of the signals (solid lines).}
\end{figure*}

\section*{7. Diffusion constant for various temperatures and different heating powers}

We compare data presented in the Fig.\,\ref{fig:Results} (in the main text) with simple free-particle diffusion model presented in Fig.\,\ref{fig:Sample} (in the main text). The model is fitted to the onset of each experimental profile yielding value of diffusion constant $D=100$\,cm$^2$/s (see Fig.\,\ref{fig:Diffusion} and Fig.\,\ref{fig:Diffusion2}).

Remarkably, we do not see variation of $D$ with $T_{bath}$ or with $I_H$.

\section*{8. Numerical calculations and material parameters}
Numerical calculation of the heat flow equation was solved in the MATLAB PDE toolbox with electron-phonon coupling and superconducting thermal conductivity calculated numerically. The heat flow from electrons to phonons in the superconducting aluminum was calculated numerically by solving integral\cite{Maisi2013_S}:
\begin{equation*}
\begin{aligned}
\dot{q}_{ep}= & \cfrac{\Sigma_{Al}}{24 \xi (5)k^5_{B}}\int\limits_{0}^{\infty} d \epsilon \epsilon^3 \left[ n \left( \epsilon, T_S \right) - n \left( \epsilon, T_P \right) \right] \\
& \times \int\limits_{-\infty}^{\infty} dE n_S \left(E \right) n_S \left(E + \epsilon \right) \left(1 - \frac{\Delta^2}{E \left(E + \epsilon \right)} \right) \\
& \times \left[f_S \left(E \right) - f_S \left(E + \epsilon \right) \right]
\end{aligned}
\end{equation*}
where $\Delta$ is the BCS temperature dependent superconducting gap, $\Sigma_{Al} = 1.8 \cdot 10^9\, W / m^3 / K^5$ is the material constant for electron-phonon coupling in aluminum, $\xi(z)$ the Riemann zeta function, $n_S(E)$ the BCS density of states, $n(\epsilon , T_P) = \{exp[\epsilon/(k_B T_P)]-1\}^{-1}$ the Bose-Einstein distribution of the phonons at temperature $T_P$.

In Figure\,\ref{fig:parameters}a numerical calculation is plotted together with analytical solution for low temperature limit:

\begin{center}
$\dot{q}_{ep}= \sum (T_e^5 - T_{ph}^5) \cdot e^{-\Delta / k_B T_S}$, $T_e<<T_c$
\end{center}

The electron-phonon coupling for copper island follows power law:
\begin{equation*}
\begin{aligned}
\dot{q}_{ep}= \Sigma_{Cu} \left( T_e^5 - T_{ph}^5 \right)
\end{aligned}
\end{equation*}
with $\Sigma_{Cu} = 2 \cdot 10^9\, W / m^3 / K^5$.

Thermal conductivity in the superconducting state was obtained as a solution of the integral\cite{Courtois2008_S} (see Figure\,\ref{fig:parameters}b.):
\begin{equation*}
\begin{aligned}
\frac{\kappa_S (T)}{\kappa_N (T)}=\frac{3}{2\pi^2}\int_{\Delta/k_BT}^{\infty}\left(\frac{x}{cosh(x/2)}\right)^2dx
\end{aligned}
\end{equation*}
where $\kappa_S (T)$ and $\kappa_N (T)$ are the thermal conductivities in the superconducting state and in the normal state respectively. $\kappa_N$ is assumed to follow linear temperature dependence defined by the Wiedemann-Franz law, i.e. $\kappa_N (T) = \sigma T L_0$, where $L_0 = 2.44 \cdot 10^{-8}$ W$\Omega$/K$^2$ is the Lorentz number and $\sigma$ is electrical conductivity ($\sigma_{Al}= 0.5 \cdot 10^8$ S/m).

Aluminum heat capacity is determined experimentally in \cite{Philips1959_S} - Figure\,\ref{fig:parameters}c. Note: $c(T_e>T_c)=\gamma T_e$, $\gamma=135\,$JK$^{-2}$m$^{-3}$ (instead of $90.9\,$JK$^{-2}$m$^{-3}$ expected for the free electron model).

In our experimental configuration the signal measured with thermometer shows increase of temperature at the level of a few dozens of milikelvins. Nevertheless the measured temperature is result of highly nonlinear heat flows on the way between the heater and the thermometer. The temperature difference between heater and thermometer after application of the heating pulse is of order of 1\,K. It follows that modeling relies on the set of temperature-dependent functions describing heat capacity, thermal conductivity and electron-phonon coupling both for aluminum wire and copper heater. The exact fitting procedure would involve rather cumbersome adjustment of these material functions. It is in contrast with an analysis in a linear regime when only single set of numbers (material parameters) is needed for comparison.


\begin{thebibliography}{44}%
\makeatletter
\providecommand \@ifxundefined [1]{%
 \@ifx{#1\undefined}
}%
\providecommand \@ifnum [1]{%
 \ifnum #1\expandafter \@firstoftwo
 \else \expandafter \@secondoftwo
 \fi
}%
\providecommand \@ifx [1]{%
 \ifx #1\expandafter \@firstoftwo
 \else \expandafter \@secondoftwo
 \fi
}%
\providecommand \natexlab [1]{#1}%
\providecommand \enquote  [1]{``#1''}%
\providecommand \bibnamefont  [1]{#1}%
\providecommand \bibfnamefont [1]{#1}%
\providecommand \citenamefont [1]{#1}%
\providecommand \href@noop [0]{\@secondoftwo}%
\providecommand \href [0]{\begingroup \@sanitize@url \@href}%
\providecommand \@href[1]{\@@startlink{#1}\@@href}%
\providecommand \@@href[1]{\endgroup#1\@@endlink}%
\providecommand \@sanitize@url [0]{\catcode `\\12\catcode `\$12\catcode
  `\&12\catcode `\#12\catcode `\^12\catcode `\_12\catcode `\%12\relax}%
\providecommand \@@startlink[1]{}%
\providecommand \@@endlink[0]{}%
\providecommand \url  [0]{\begingroup\@sanitize@url \@url }%
\providecommand \@url [1]{\endgroup\@href {#1}{\urlprefix }}%
\providecommand \urlprefix  [0]{URL }%
\providecommand \Eprint [0]{\href }%
\providecommand \doibase [0]{http://dx.doi.org/}%
\providecommand \selectlanguage [0]{\@gobble}%
\providecommand \bibinfo  [0]{\@secondoftwo}%
\providecommand \bibfield  [0]{\@secondoftwo}%
\providecommand \translation [1]{[#1]}%
\providecommand \BibitemOpen [0]{}%
\providecommand \bibitemStop [0]{}%
\providecommand \bibitemNoStop [0]{.\EOS\space}%
\providecommand \EOS [0]{\spacefactor3000\relax}%
\providecommand \BibitemShut  [1]{\csname bibitem#1\endcsname}%
\let\auto@bib@innerbib\@empty
\bibitem [{\citenamefont {Wellstood}\ \emph {et~al.}(1994)\citenamefont
  {Wellstood}, \citenamefont {Urbina},\ and\ \citenamefont
  {Clarke}}]{Clarke1994}%
  \BibitemOpen
  \bibfield  {author} {\bibinfo {author} {\bibfnamefont {F.~C.}\ \bibnamefont
  {Wellstood}}, \bibinfo {author} {\bibfnamefont {C.}~\bibnamefont {Urbina}}, \
  and\ \bibinfo {author} {\bibfnamefont {J.}~\bibnamefont {Clarke}},\ }\href
  {\doibase 10.1103/PhysRevB.49.5942} {\bibfield  {journal} {\bibinfo
  {journal} {Phys. Rev. B}\ }\textbf {\bibinfo {volume} {49}},\ \bibinfo
  {pages} {5942} (\bibinfo {year} {1994})}\BibitemShut {NoStop}%
\bibitem [{\citenamefont {Timofeev}\ \emph {et~al.}(2009)\citenamefont
  {Timofeev}, \citenamefont {Garc\'{\i}a}, \citenamefont {Kopnin},
  \citenamefont {Savin}, \citenamefont {Meschke}, \citenamefont {Giazotto},\
  and\ \citenamefont {Pekola}}]{Pekola2009}%
  \BibitemOpen
  \bibfield  {author} {\bibinfo {author} {\bibfnamefont {A.~V.}\ \bibnamefont
  {Timofeev}}, \bibinfo {author} {\bibfnamefont {C.~P.}\ \bibnamefont
  {Garc\'{\i}a}}, \bibinfo {author} {\bibfnamefont {N.~B.}\ \bibnamefont
  {Kopnin}}, \bibinfo {author} {\bibfnamefont {A.~M.}\ \bibnamefont {Savin}},
  \bibinfo {author} {\bibfnamefont {M.}~\bibnamefont {Meschke}}, \bibinfo
  {author} {\bibfnamefont {F.}~\bibnamefont {Giazotto}}, \ and\ \bibinfo
  {author} {\bibfnamefont {J.~P.}\ \bibnamefont {Pekola}},\ }\href {\doibase
  10.1103/PhysRevLett.102.017003} {\bibfield  {journal} {\bibinfo  {journal}
  {Phys. Rev. Lett.}\ }\textbf {\bibinfo {volume} {102}},\ \bibinfo {pages}
  {017003} (\bibinfo {year} {2009})}\BibitemShut {NoStop}%
\bibitem [{\citenamefont {Meschke}\ \emph {et~al.}(2006)\citenamefont
  {Meschke}, \citenamefont {Guichard},\ and\ \citenamefont
  {Pekola}}]{Pekola2006}%
  \BibitemOpen
  \bibfield  {author} {\bibinfo {author} {\bibfnamefont {M.}~\bibnamefont
  {Meschke}}, \bibinfo {author} {\bibfnamefont {W.}~\bibnamefont {Guichard}}, \
  and\ \bibinfo {author} {\bibfnamefont {J.~P.}\ \bibnamefont {Pekola}},\
  }\href {\doibase 10.1038/nature05276} {\bibfield  {journal} {\bibinfo
  {journal} {Nature}\ }\textbf {\bibinfo {volume} {444}},\ \bibinfo {pages}
  {187} (\bibinfo {year} {2006})}\BibitemShut {NoStop}%
\bibitem [{\citenamefont {Ullom}\ \emph {et~al.}(1998)\citenamefont {Ullom},
  \citenamefont {Fisher},\ and\ \citenamefont {Nahum}}]{Ullom1998}%
  \BibitemOpen
  \bibfield  {author} {\bibinfo {author} {\bibfnamefont {J.~N.}\ \bibnamefont
  {Ullom}}, \bibinfo {author} {\bibfnamefont {P.~A.}\ \bibnamefont {Fisher}}, \
  and\ \bibinfo {author} {\bibfnamefont {M.}~\bibnamefont {Nahum}},\ }\href
  {\doibase 10.1103/PhysRevB.58.8225} {\bibfield  {journal} {\bibinfo
  {journal} {Phys. Rev. B}\ }\textbf {\bibinfo {volume} {58}},\ \bibinfo
  {pages} {8225} (\bibinfo {year} {1998})}\BibitemShut {NoStop}%
\bibitem [{\citenamefont {Courtois}\ \emph {et~al.}(2008)\citenamefont
  {Courtois}, \citenamefont {Meschke}, \citenamefont {Peltonen},\ and\
  \citenamefont {Pekola}}]{Pekola2008}%
  \BibitemOpen
  \bibfield  {author} {\bibinfo {author} {\bibfnamefont {H.}~\bibnamefont
  {Courtois}}, \bibinfo {author} {\bibfnamefont {M.}~\bibnamefont {Meschke}},
  \bibinfo {author} {\bibfnamefont {J.~T.}\ \bibnamefont {Peltonen}}, \ and\
  \bibinfo {author} {\bibfnamefont {J.~P.}\ \bibnamefont {Pekola}},\ }\href
  {\doibase 10.1103/PhysRevLett.101.067002} {\bibfield  {journal} {\bibinfo
  {journal} {Phys. Rev. Lett.}\ }\textbf {\bibinfo {volume} {101}},\ \bibinfo
  {pages} {067002} (\bibinfo {year} {2008})}\BibitemShut {NoStop}%
\bibitem [{\citenamefont {Giazotto}\ and\ \citenamefont
  {Mart\'{\i}nez-P{\'e}rez}(2012)}]{Giazotto2012}%
  \BibitemOpen
  \bibfield  {author} {\bibinfo {author} {\bibfnamefont {F.}~\bibnamefont
  {Giazotto}}\ and\ \bibinfo {author} {\bibfnamefont {M.~J.}\ \bibnamefont
  {Mart\'{\i}nez-P{\'e}rez}},\ }\href {\doibase 10.1038/nature11702} {\bibfield
   {journal} {\bibinfo  {journal} {Nature}\ }\textbf {\bibinfo {volume}
  {492}},\ \bibinfo {pages} {401} (\bibinfo {year} {2012})}\BibitemShut
  {NoStop}%
\bibitem [{\citenamefont {Jezouin}\ \emph {et~al.}(2013)\citenamefont
  {Jezouin}, \citenamefont {Parmentier}, \citenamefont {Anthore}, \citenamefont
  {Gennser}, \citenamefont {Cavanna}, \citenamefont {Jin},\ and\ \citenamefont
  {Pierre}}]{Jezouin2013}%
  \BibitemOpen
  \bibfield  {author} {\bibinfo {author} {\bibfnamefont {S.}~\bibnamefont
  {Jezouin}}, \bibinfo {author} {\bibfnamefont {F.~D.}\ \bibnamefont
  {Parmentier}}, \bibinfo {author} {\bibfnamefont {A.}~\bibnamefont {Anthore}},
  \bibinfo {author} {\bibfnamefont {U.}~\bibnamefont {Gennser}}, \bibinfo
  {author} {\bibfnamefont {A.}~\bibnamefont {Cavanna}}, \bibinfo {author}
  {\bibfnamefont {Y.}~\bibnamefont {Jin}}, \ and\ \bibinfo {author}
  {\bibfnamefont {F.}~\bibnamefont {Pierre}},\ }\href {\doibase
  10.1126/science.1241912} {\bibfield  {journal} {\bibinfo  {journal}
  {Science}\ }\textbf {\bibinfo {volume} {342}},\ \bibinfo {pages} {601}
  (\bibinfo {year} {2013})}\BibitemShut {NoStop}%
\bibitem [{\citenamefont {Gasparinetti}\ \emph {et~al.}(2011)\citenamefont
  {Gasparinetti}, \citenamefont {Deon}, \citenamefont {Biasiol}, \citenamefont
  {Sorba}, \citenamefont {Beltram},\ and\ \citenamefont
  {Giazotto}}]{Gasparinetti2011}%
  \BibitemOpen
  \bibfield  {author} {\bibinfo {author} {\bibfnamefont {S.}~\bibnamefont
  {Gasparinetti}}, \bibinfo {author} {\bibfnamefont {F.}~\bibnamefont {Deon}},
  \bibinfo {author} {\bibfnamefont {G.}~\bibnamefont {Biasiol}}, \bibinfo
  {author} {\bibfnamefont {L.}~\bibnamefont {Sorba}}, \bibinfo {author}
  {\bibfnamefont {F.}~\bibnamefont {Beltram}}, \ and\ \bibinfo {author}
  {\bibfnamefont {F.}~\bibnamefont {Giazotto}},\ }\href {\doibase
  10.1103/PhysRevB.83.201306} {\bibfield  {journal} {\bibinfo  {journal} {Phys.
  Rev. B}\ }\textbf {\bibinfo {volume} {83}},\ \bibinfo {pages} {201306}
  (\bibinfo {year} {2011})}\BibitemShut {NoStop}%
\bibitem [{\citenamefont {Betz}\ \emph {et~al.}(2012)\citenamefont {Betz},
  \citenamefont {Vialla}, \citenamefont {Brunel}, \citenamefont {Voisin},
  \citenamefont {Picher}, \citenamefont {Cavanna}, \citenamefont {Madouri},
  \citenamefont {F\`eve}, \citenamefont {Berroir}, \citenamefont
  {Pla\ifmmode~\mbox{\c{c}}\else \c{c}\fi{}ais},\ and\ \citenamefont
  {Pallecchi}}]{Betz2012}%
  \BibitemOpen
  \bibfield  {author} {\bibinfo {author} {\bibfnamefont {A.~C.}\ \bibnamefont
  {Betz}}, \bibinfo {author} {\bibfnamefont {F.}~\bibnamefont {Vialla}},
  \bibinfo {author} {\bibfnamefont {D.}~\bibnamefont {Brunel}}, \bibinfo
  {author} {\bibfnamefont {C.}~\bibnamefont {Voisin}}, \bibinfo {author}
  {\bibfnamefont {M.}~\bibnamefont {Picher}}, \bibinfo {author} {\bibfnamefont
  {A.}~\bibnamefont {Cavanna}}, \bibinfo {author} {\bibfnamefont
  {A.}~\bibnamefont {Madouri}}, \bibinfo {author} {\bibfnamefont
  {G.}~\bibnamefont {F\`eve}}, \bibinfo {author} {\bibfnamefont {J.-M.}\
  \bibnamefont {Berroir}}, \bibinfo {author} {\bibfnamefont {B.}~\bibnamefont
  {Pla\ifmmode~\mbox{\c{c}}\else \c{c}\fi{}ais}}, \ and\ \bibinfo {author}
  {\bibfnamefont {E.}~\bibnamefont {Pallecchi}},\ }\href {\doibase
  10.1103/PhysRevLett.109.056805} {\bibfield  {journal} {\bibinfo  {journal}
  {Phys. Rev. Lett.}\ }\textbf {\bibinfo {volume} {109}},\ \bibinfo {pages}
  {056805} (\bibinfo {year} {2012})}\BibitemShut {NoStop}%
\bibitem [{\citenamefont {Michon}\ \emph {et~al.}(2018)\citenamefont {Michon},
  \citenamefont {Ataei}, \citenamefont {Bourgeois-Hope}, \citenamefont
  {Collignon}, \citenamefont {Li}, \citenamefont {Badoux}, \citenamefont
  {Gourgout}, \citenamefont {Lalibert\'e}, \citenamefont {Zhou}, \citenamefont
  {Doiron-Leyraud},\ and\ \citenamefont {Taillefer}}]{Michon2018}%
  \BibitemOpen
  \bibfield  {author} {\bibinfo {author} {\bibfnamefont {B.}~\bibnamefont
  {Michon}}, \bibinfo {author} {\bibfnamefont {A.}~\bibnamefont {Ataei}},
  \bibinfo {author} {\bibfnamefont {P.}~\bibnamefont {Bourgeois-Hope}},
  \bibinfo {author} {\bibfnamefont {C.}~\bibnamefont {Collignon}}, \bibinfo
  {author} {\bibfnamefont {S.~Y.}\ \bibnamefont {Li}}, \bibinfo {author}
  {\bibfnamefont {S.}~\bibnamefont {Badoux}}, \bibinfo {author} {\bibfnamefont
  {A.}~\bibnamefont {Gourgout}}, \bibinfo {author} {\bibfnamefont
  {F.}~\bibnamefont {Lalibert\'e}}, \bibinfo {author} {\bibfnamefont {J.-S.}\
  \bibnamefont {Zhou}}, \bibinfo {author} {\bibfnamefont {N.}~\bibnamefont
  {Doiron-Leyraud}}, \ and\ \bibinfo {author} {\bibfnamefont {L.}~\bibnamefont
  {Taillefer}},\ }\href {\doibase 10.1103/PhysRevX.8.041010} {\bibfield
  {journal} {\bibinfo  {journal} {Phys. Rev. X}\ }\textbf {\bibinfo {volume}
  {8}},\ \bibinfo {pages} {041010} (\bibinfo {year} {2018})}\BibitemShut
  {NoStop}%
\bibitem [{\citenamefont {Schmidt}\ \emph {et~al.}(2004)\citenamefont
  {Schmidt}, \citenamefont {Yung},\ and\ \citenamefont
  {Cleland}}]{Cleland2004}%
  \BibitemOpen
  \bibfield  {author} {\bibinfo {author} {\bibfnamefont {D.~R.}\ \bibnamefont
  {Schmidt}}, \bibinfo {author} {\bibfnamefont {C.~S.}\ \bibnamefont {Yung}}, \
  and\ \bibinfo {author} {\bibfnamefont {A.~N.}\ \bibnamefont {Cleland}},\
  }\href {\doibase 10.1103/PhysRevB.69.140301} {\bibfield  {journal} {\bibinfo
  {journal} {Phys. Rev. B}\ }\textbf {\bibinfo {volume} {69}},\ \bibinfo
  {pages} {140301} (\bibinfo {year} {2004})}\BibitemShut {NoStop}%
\bibitem [{\citenamefont {Saira}\ \emph {et~al.}(2016)\citenamefont {Saira},
  \citenamefont {Zgirski}, \citenamefont {Viisanen}, \citenamefont {Golubev},\
  and\ \citenamefont {Pekola}}]{Zgirski2016}%
  \BibitemOpen
  \bibfield  {author} {\bibinfo {author} {\bibfnamefont {O.-P.}\ \bibnamefont
  {Saira}}, \bibinfo {author} {\bibfnamefont {M.}~\bibnamefont {Zgirski}},
  \bibinfo {author} {\bibfnamefont {K.~L.}\ \bibnamefont {Viisanen}}, \bibinfo
  {author} {\bibfnamefont {D.~S.}\ \bibnamefont {Golubev}}, \ and\ \bibinfo
  {author} {\bibfnamefont {J.~P.}\ \bibnamefont {Pekola}},\ }\href {\doibase
  10.1103/PhysRevApplied.6.024005} {\bibfield  {journal} {\bibinfo  {journal}
  {Phys. Rev. Applied}\ }\textbf {\bibinfo {volume} {6}},\ \bibinfo {pages}
  {024005} (\bibinfo {year} {2016})}\BibitemShut {NoStop}%
\bibitem [{\citenamefont {Gasparinetti}\ \emph {et~al.}(2015)\citenamefont
  {Gasparinetti}, \citenamefont {Viisanen}, \citenamefont {Saira},
  \citenamefont {Faivre}, \citenamefont {Arzeo}, \citenamefont {Meschke},\ and\
  \citenamefont {Pekola}}]{Gasparinetti2015}%
  \BibitemOpen
  \bibfield  {author} {\bibinfo {author} {\bibfnamefont {S.}~\bibnamefont
  {Gasparinetti}}, \bibinfo {author} {\bibfnamefont {K.~L.}\ \bibnamefont
  {Viisanen}}, \bibinfo {author} {\bibfnamefont {O.-P.}\ \bibnamefont {Saira}},
  \bibinfo {author} {\bibfnamefont {T.}~\bibnamefont {Faivre}}, \bibinfo
  {author} {\bibfnamefont {M.}~\bibnamefont {Arzeo}}, \bibinfo {author}
  {\bibfnamefont {M.}~\bibnamefont {Meschke}}, \ and\ \bibinfo {author}
  {\bibfnamefont {J.~P.}\ \bibnamefont {Pekola}},\ }\href {\doibase
  10.1103/PhysRevApplied.3.014007} {\bibfield  {journal} {\bibinfo  {journal}
  {Phys. Rev. Applied}\ }\textbf {\bibinfo {volume} {3}},\ \bibinfo {pages}
  {014007} (\bibinfo {year} {2015})}\BibitemShut {NoStop}%
\bibitem [{\citenamefont {Barends}\ \emph {et~al.}(2008)\citenamefont
  {Barends}, \citenamefont {Baselmans}, \citenamefont {Yates}, \citenamefont
  {Gao}, \citenamefont {Hovenier},\ and\ \citenamefont
  {Klapwijk}}]{Klapwijk2008}%
  \BibitemOpen
  \bibfield  {author} {\bibinfo {author} {\bibfnamefont {R.}~\bibnamefont
  {Barends}}, \bibinfo {author} {\bibfnamefont {J.~J.~A.}\ \bibnamefont
  {Baselmans}}, \bibinfo {author} {\bibfnamefont {S.~J.~C.}\ \bibnamefont
  {Yates}}, \bibinfo {author} {\bibfnamefont {J.~R.}\ \bibnamefont {Gao}},
  \bibinfo {author} {\bibfnamefont {J.~N.}\ \bibnamefont {Hovenier}}, \ and\
  \bibinfo {author} {\bibfnamefont {T.~M.}\ \bibnamefont {Klapwijk}},\ }\href
  {\doibase 10.1103/PhysRevLett.100.257002} {\bibfield  {journal} {\bibinfo
  {journal} {Phys. Rev. Lett.}\ }\textbf {\bibinfo {volume} {100}},\ \bibinfo
  {pages} {257002} (\bibinfo {year} {2008})}\BibitemShut {NoStop}%
\bibitem [{\citenamefont {Viisanen}\ and\ \citenamefont
  {Pekola}(2018)}]{Pekola2018}%
  \BibitemOpen
  \bibfield  {author} {\bibinfo {author} {\bibfnamefont {K.~L.}\ \bibnamefont
  {Viisanen}}\ and\ \bibinfo {author} {\bibfnamefont {J.~P.}\ \bibnamefont
  {Pekola}},\ }\href {\doibase 10.1103/PhysRevB.97.115422} {\bibfield
  {journal} {\bibinfo  {journal} {Phys. Rev. B}\ }\textbf {\bibinfo {volume}
  {97}},\ \bibinfo {pages} {115422} (\bibinfo {year} {2018})}\BibitemShut
  {NoStop}%
\bibitem [{\citenamefont {Zgirski}\ \emph {et~al.}(2018)\citenamefont
  {Zgirski}, \citenamefont {Foltyn}, \citenamefont {Savin}, \citenamefont
  {Norowski}, \citenamefont {Meschke},\ and\ \citenamefont
  {Pekola}}]{Zgirski2018}%
  \BibitemOpen
  \bibfield  {author} {\bibinfo {author} {\bibfnamefont {M.}~\bibnamefont
  {Zgirski}}, \bibinfo {author} {\bibfnamefont {M.}~\bibnamefont {Foltyn}},
  \bibinfo {author} {\bibfnamefont {A.}~\bibnamefont {Savin}}, \bibinfo
  {author} {\bibfnamefont {K.}~\bibnamefont {Norowski}}, \bibinfo {author}
  {\bibfnamefont {M.}~\bibnamefont {Meschke}}, \ and\ \bibinfo {author}
  {\bibfnamefont {J.}~\bibnamefont {Pekola}},\ }\href {\doibase
  10.1103/PhysRevApplied.10.044068} {\bibfield  {journal} {\bibinfo  {journal}
  {Phys. Rev. Applied}\ }\textbf {\bibinfo {volume} {10}},\ \bibinfo {pages}
  {044068} (\bibinfo {year} {2018})}\BibitemShut {NoStop}%
\bibitem [{\citenamefont {Wei}\ \emph {et~al.}(2008)\citenamefont {Wei},
  \citenamefont {Olaya}, \citenamefont {Karasik}, \citenamefont {Pereverzev},
  \citenamefont {Sergeev},\ and\ \citenamefont {Gershenson}}]{Wei2008}%
  \BibitemOpen
  \bibfield  {author} {\bibinfo {author} {\bibfnamefont {J.}~\bibnamefont
  {Wei}}, \bibinfo {author} {\bibfnamefont {D.}~\bibnamefont {Olaya}}, \bibinfo
  {author} {\bibfnamefont {B.~S.}\ \bibnamefont {Karasik}}, \bibinfo {author}
  {\bibfnamefont {S.~V.}\ \bibnamefont {Pereverzev}}, \bibinfo {author}
  {\bibfnamefont {A.~V.}\ \bibnamefont {Sergeev}}, \ and\ \bibinfo {author}
  {\bibfnamefont {M.~E.}\ \bibnamefont {Gershenson}},\ }\href {\doibase
  10.1038/nnano.2008.173} {\bibfield  {journal} {\bibinfo  {journal} {Nat.
  Nanotechnol.}\ }\textbf {\bibinfo {volume} {3}},\ \bibinfo {pages} {496}
  (\bibinfo {year} {2008})}\BibitemShut {NoStop}%
\bibitem [{\citenamefont {Govenius}\ \emph {et~al.}(2016)\citenamefont
  {Govenius}, \citenamefont {Lake}, \citenamefont {Tan},\ and\ \citenamefont
  {M\"ott\"onen}}]{Govenius2016}%
  \BibitemOpen
  \bibfield  {author} {\bibinfo {author} {\bibfnamefont {J.}~\bibnamefont
  {Govenius}}, \bibinfo {author} {\bibfnamefont {R.~E.}\ \bibnamefont {Lake}},
  \bibinfo {author} {\bibfnamefont {K.~Y.}\ \bibnamefont {Tan}}, \ and\
  \bibinfo {author} {\bibfnamefont {M.}~\bibnamefont {M\"ott\"onen}},\ }\href
  {\doibase 10.1103/PhysRevLett.117.030802} {\bibfield  {journal} {\bibinfo
  {journal} {Phys. Rev. Lett.}\ }\textbf {\bibinfo {volume} {117}},\ \bibinfo
  {pages} {030802} (\bibinfo {year} {2016})}\BibitemShut {NoStop}%
\bibitem [{\citenamefont {Walsh}\ \emph {et~al.}(2017)\citenamefont {Walsh},
  \citenamefont {Efetov}, \citenamefont {Lee}, \citenamefont {Heuck},
  \citenamefont {Crossno}, \citenamefont {Ohki}, \citenamefont {Kim},
  \citenamefont {Englund},\ and\ \citenamefont {Fong}}]{Walsh2017}%
  \BibitemOpen
  \bibfield  {author} {\bibinfo {author} {\bibfnamefont {E.~D.}\ \bibnamefont
  {Walsh}}, \bibinfo {author} {\bibfnamefont {D.~K.}\ \bibnamefont {Efetov}},
  \bibinfo {author} {\bibfnamefont {G.-H.}\ \bibnamefont {Lee}}, \bibinfo
  {author} {\bibfnamefont {M.}~\bibnamefont {Heuck}}, \bibinfo {author}
  {\bibfnamefont {J.}~\bibnamefont {Crossno}}, \bibinfo {author} {\bibfnamefont
  {T.~A.}\ \bibnamefont {Ohki}}, \bibinfo {author} {\bibfnamefont
  {P.}~\bibnamefont {Kim}}, \bibinfo {author} {\bibfnamefont {D.}~\bibnamefont
  {Englund}}, \ and\ \bibinfo {author} {\bibfnamefont {K.~C.}\ \bibnamefont
  {Fong}},\ }\href {\doibase 10.1103/PhysRevApplied.8.024022} {\bibfield
  {journal} {\bibinfo  {journal} {Phys. Rev. Applied}\ }\textbf {\bibinfo
  {volume} {8}},\ \bibinfo {pages} {024022} (\bibinfo {year}
  {2017})}\BibitemShut {NoStop}%
\bibitem [{\citenamefont {Giazotto}\ \emph {et~al.}(2006)\citenamefont
  {Giazotto}, \citenamefont {Heikkil\"a}, \citenamefont {Luukanen},
  \citenamefont {Savin},\ and\ \citenamefont {Pekola}}]{Savin2006}%
  \BibitemOpen
  \bibfield  {author} {\bibinfo {author} {\bibfnamefont {F.}~\bibnamefont
  {Giazotto}}, \bibinfo {author} {\bibfnamefont {T.~T.}\ \bibnamefont
  {Heikkil\"a}}, \bibinfo {author} {\bibfnamefont {A.}~\bibnamefont
  {Luukanen}}, \bibinfo {author} {\bibfnamefont {A.~M.}\ \bibnamefont {Savin}},
  \ and\ \bibinfo {author} {\bibfnamefont {J.~P.}\ \bibnamefont {Pekola}},\
  }\href {\doibase 10.1103/RevModPhys.78.217} {\bibfield  {journal} {\bibinfo
  {journal} {Rev. Mod. Phys.}\ }\textbf {\bibinfo {volume} {78}},\ \bibinfo
  {pages} {217} (\bibinfo {year} {2006})}\BibitemShut {NoStop}%
\bibitem [{\citenamefont {Serniak}\ \emph {et~al.}(2018)\citenamefont
  {Serniak}, \citenamefont {Hays}, \citenamefont {de~Lange}, \citenamefont
  {Diamond}, \citenamefont {Shankar}, \citenamefont {Burkhart}, \citenamefont
  {Frunzio}, \citenamefont {Houzet},\ and\ \citenamefont
  {Devoret}}]{Serniak2018}%
  \BibitemOpen
  \bibfield  {author} {\bibinfo {author} {\bibfnamefont {K.}~\bibnamefont
  {Serniak}}, \bibinfo {author} {\bibfnamefont {M.}~\bibnamefont {Hays}},
  \bibinfo {author} {\bibfnamefont {G.}~\bibnamefont {de~Lange}}, \bibinfo
  {author} {\bibfnamefont {S.}~\bibnamefont {Diamond}}, \bibinfo {author}
  {\bibfnamefont {S.}~\bibnamefont {Shankar}}, \bibinfo {author} {\bibfnamefont
  {L.~D.}\ \bibnamefont {Burkhart}}, \bibinfo {author} {\bibfnamefont
  {L.}~\bibnamefont {Frunzio}}, \bibinfo {author} {\bibfnamefont
  {M.}~\bibnamefont {Houzet}}, \ and\ \bibinfo {author} {\bibfnamefont {M.~H.}\
  \bibnamefont {Devoret}},\ }\href {\doibase 10.1103/PhysRevLett.121.157701}
  {\bibfield  {journal} {\bibinfo  {journal} {Phys. Rev. Lett.}\ }\textbf
  {\bibinfo {volume} {121}},\ \bibinfo {pages} {157701} (\bibinfo {year}
  {2018})}\BibitemShut {NoStop}%
\bibitem [{\citenamefont {Rainis}\ and\ \citenamefont
  {Loss}(2012)}]{Rainis2012}%
  \BibitemOpen
  \bibfield  {author} {\bibinfo {author} {\bibfnamefont {D.}~\bibnamefont
  {Rainis}}\ and\ \bibinfo {author} {\bibfnamefont {D.}~\bibnamefont {Loss}},\
  }\href {\doibase 10.1103/PhysRevB.85.174533} {\bibfield  {journal} {\bibinfo
  {journal} {Phys. Rev. B}\ }\textbf {\bibinfo {volume} {85}},\ \bibinfo
  {pages} {174533} (\bibinfo {year} {2012})}\BibitemShut {NoStop}%
\bibitem [{\citenamefont {Aasen}\ \emph {et~al.}(2016)\citenamefont {Aasen},
  \citenamefont {Hell}, \citenamefont {Mishmash}, \citenamefont {Higginbotham},
  \citenamefont {Danon}, \citenamefont {Leijnse}, \citenamefont {Jespersen},
  \citenamefont {Folk}, \citenamefont {Marcus}, \citenamefont {Flensberg},\
  and\ \citenamefont {Alicea}}]{Aasen2016}%
  \BibitemOpen
  \bibfield  {author} {\bibinfo {author} {\bibfnamefont {D.}~\bibnamefont
  {Aasen}}, \bibinfo {author} {\bibfnamefont {M.}~\bibnamefont {Hell}},
  \bibinfo {author} {\bibfnamefont {R.~V.}\ \bibnamefont {Mishmash}}, \bibinfo
  {author} {\bibfnamefont {A.}~\bibnamefont {Higginbotham}}, \bibinfo {author}
  {\bibfnamefont {J.}~\bibnamefont {Danon}}, \bibinfo {author} {\bibfnamefont
  {M.}~\bibnamefont {Leijnse}}, \bibinfo {author} {\bibfnamefont {T.~S.}\
  \bibnamefont {Jespersen}}, \bibinfo {author} {\bibfnamefont {J.~A.}\
  \bibnamefont {Folk}}, \bibinfo {author} {\bibfnamefont {C.~M.}\ \bibnamefont
  {Marcus}}, \bibinfo {author} {\bibfnamefont {K.}~\bibnamefont {Flensberg}}, \
  and\ \bibinfo {author} {\bibfnamefont {J.}~\bibnamefont {Alicea}},\ }\href
  {\doibase 10.1103/PhysRevX.6.031016} {\bibfield  {journal} {\bibinfo
  {journal} {Phys. Rev. X}\ }\textbf {\bibinfo {volume} {6}},\ \bibinfo {pages}
  {031016} (\bibinfo {year} {2016})}\BibitemShut {NoStop}%
\bibitem [{\citenamefont {Pekola}(2015)}]{Pekola2015}%
  \BibitemOpen
  \bibfield  {author} {\bibinfo {author} {\bibfnamefont {J.~P.}\ \bibnamefont
  {Pekola}},\ }\href {\doibase 10.1038/nphys3169} {\bibfield  {journal}
  {\bibinfo  {journal} {Nat. Phys.}\ }\textbf {\bibinfo {volume} {11}},\
  \bibinfo {pages} {118} (\bibinfo {year} {2015})}\BibitemShut {NoStop}%
\bibitem [{\citenamefont {Fornieri}\ and\ \citenamefont
  {Giazotto}(2017)}]{Fornieri2017}%
  \BibitemOpen
  \bibfield  {author} {\bibinfo {author} {\bibfnamefont {A.}~\bibnamefont
  {Fornieri}}\ and\ \bibinfo {author} {\bibfnamefont {F.}~\bibnamefont
  {Giazotto}},\ }\href {\doibase 10.1038/nnano.2017.204} {\bibfield  {journal}
  {\bibinfo  {journal} {Nat. Nanotechnol.}\ }\textbf {\bibinfo {volume} {12}},\
  \bibinfo {pages} {944} (\bibinfo {year} {2017})}\BibitemShut {NoStop}%
\bibitem [{\citenamefont {Fornieri}\ \emph {et~al.}(2017)\citenamefont
  {Fornieri}, \citenamefont {Timossi}, \citenamefont {Virtanen}, \citenamefont
  {Solinas},\ and\ \citenamefont {Giazotto}}]{Giazotto2017}%
  \BibitemOpen
  \bibfield  {author} {\bibinfo {author} {\bibfnamefont {A.}~\bibnamefont
  {Fornieri}}, \bibinfo {author} {\bibfnamefont {G.}~\bibnamefont {Timossi}},
  \bibinfo {author} {\bibfnamefont {P.}~\bibnamefont {Virtanen}}, \bibinfo
  {author} {\bibfnamefont {P.}~\bibnamefont {Solinas}}, \ and\ \bibinfo
  {author} {\bibfnamefont {F.}~\bibnamefont {Giazotto}},\ }\href {\doibase
  10.1038/nnano.2017.25} {\bibfield  {journal} {\bibinfo  {journal} {Nat.
  Nanotechnol.}\ }\textbf {\bibinfo {volume} {12}},\ \bibinfo {pages} {425}
  (\bibinfo {year} {2017})}\BibitemShut {NoStop}%
\bibitem [{\citenamefont {Mart{\'i}nez-P{\'e}rez}\ \emph
  {et~al.}(2014)\citenamefont {Mart{\'i}nez-P{\'e}rez}, \citenamefont
  {Solinas},\ and\ \citenamefont {Giazotto}}]{Giazotto2014}%
  \BibitemOpen
  \bibfield  {author} {\bibinfo {author} {\bibfnamefont {M.~J.}\ \bibnamefont
  {Mart{\'i}nez-P{\'e}rez}}, \bibinfo {author} {\bibfnamefont {P.}~\bibnamefont
  {Solinas}}, \ and\ \bibinfo {author} {\bibfnamefont {F.}~\bibnamefont
  {Giazotto}},\ }\href {\doibase 10.1007/s10909-014-1132-6} {\bibfield
  {journal} {\bibinfo  {journal} {J Low Temp Phys}\ }\textbf {\bibinfo {volume}
  {175}},\ \bibinfo {pages} {813} (\bibinfo {year} {2014})}\BibitemShut
  {NoStop}%
\bibitem [{\citenamefont {Li}\ \emph {et~al.}(2012)\citenamefont {Li},
  \citenamefont {Ren}, \citenamefont {Wang}, \citenamefont {Zhang},
  \citenamefont {H\"anggi},\ and\ \citenamefont {Li}}]{Li2012}%
  \BibitemOpen
  \bibfield  {author} {\bibinfo {author} {\bibfnamefont {N.}~\bibnamefont
  {Li}}, \bibinfo {author} {\bibfnamefont {J.}~\bibnamefont {Ren}}, \bibinfo
  {author} {\bibfnamefont {L.}~\bibnamefont {Wang}}, \bibinfo {author}
  {\bibfnamefont {G.}~\bibnamefont {Zhang}}, \bibinfo {author} {\bibfnamefont
  {P.}~\bibnamefont {H\"anggi}}, \ and\ \bibinfo {author} {\bibfnamefont
  {B.}~\bibnamefont {Li}},\ }\href {\doibase 10.1103/RevModPhys.84.1045}
  {\bibfield  {journal} {\bibinfo  {journal} {Rev. Mod. Phys.}\ }\textbf
  {\bibinfo {volume} {84}},\ \bibinfo {pages} {1045} (\bibinfo {year}
  {2012})}\BibitemShut {NoStop}%
\bibitem [{\citenamefont {Bauer}\ \emph {et~al.}(2012)\citenamefont {Bauer},
  \citenamefont {Saitoh},\ and\ \citenamefont {van Wees}}]{Bauer2012}%
  \BibitemOpen
  \bibfield  {author} {\bibinfo {author} {\bibfnamefont {G.~E.~W.}\
  \bibnamefont {Bauer}}, \bibinfo {author} {\bibfnamefont {E.}~\bibnamefont
  {Saitoh}}, \ and\ \bibinfo {author} {\bibfnamefont {B.~J.}\ \bibnamefont {van
  Wees}},\ }\href {\doibase 10.1038/nmat3301} {\bibfield  {journal} {\bibinfo
  {journal} {Nat. Mater}\ }\textbf {\bibinfo {volume} {11}},\ \bibinfo {pages}
  {391} (\bibinfo {year} {2012})}\BibitemShut {NoStop}%
\bibitem [{\citenamefont {Roukes}\ \emph {et~al.}(1985)\citenamefont {Roukes},
  \citenamefont {Freeman}, \citenamefont {Germain}, \citenamefont
  {Richardson},\ and\ \citenamefont {Ketchen}}]{Roukes1985}%
  \BibitemOpen
  \bibfield  {author} {\bibinfo {author} {\bibfnamefont {M.~L.}\ \bibnamefont
  {Roukes}}, \bibinfo {author} {\bibfnamefont {M.~R.}\ \bibnamefont {Freeman}},
  \bibinfo {author} {\bibfnamefont {R.~S.}\ \bibnamefont {Germain}}, \bibinfo
  {author} {\bibfnamefont {R.~C.}\ \bibnamefont {Richardson}}, \ and\ \bibinfo
  {author} {\bibfnamefont {M.~B.}\ \bibnamefont {Ketchen}},\ }\href {\doibase
  10.1103/PhysRevLett.55.422} {\bibfield  {journal} {\bibinfo  {journal} {Phys.
  Rev. Lett.}\ }\textbf {\bibinfo {volume} {55}},\ \bibinfo {pages} {422}
  (\bibinfo {year} {1985})}\BibitemShut {NoStop}%
\bibitem [{\citenamefont {Gershenson}\ \emph {et~al.}(2001)\citenamefont
  {Gershenson}, \citenamefont {Gong}, \citenamefont {Sato}, \citenamefont
  {Karasik},\ and\ \citenamefont {Sergeev}}]{Gershenson2001}%
  \BibitemOpen
  \bibfield  {author} {\bibinfo {author} {\bibfnamefont {M.~E.}\ \bibnamefont
  {Gershenson}}, \bibinfo {author} {\bibfnamefont {D.}~\bibnamefont {Gong}},
  \bibinfo {author} {\bibfnamefont {T.}~\bibnamefont {Sato}}, \bibinfo {author}
  {\bibfnamefont {B.~S.}\ \bibnamefont {Karasik}}, \ and\ \bibinfo {author}
  {\bibfnamefont {A.~V.}\ \bibnamefont {Sergeev}},\ }\href {\doibase
  10.1063/1.1407302} {\bibfield  {journal} {\bibinfo  {journal} {Applied
  Physics Letters}\ }\textbf {\bibinfo {volume} {79}},\ \bibinfo {pages} {2049}
  (\bibinfo {year} {2001})}\BibitemShut {NoStop}%
\bibitem [{\citenamefont {Huard}\ \emph {et~al.}(2007)\citenamefont {Huard},
  \citenamefont {Pothier}, \citenamefont {Esteve},\ and\ \citenamefont
  {Nagaev}}]{Huard2007}%
  \BibitemOpen
  \bibfield  {author} {\bibinfo {author} {\bibfnamefont {B.}~\bibnamefont
  {Huard}}, \bibinfo {author} {\bibfnamefont {H.}~\bibnamefont {Pothier}},
  \bibinfo {author} {\bibfnamefont {D.}~\bibnamefont {Esteve}}, \ and\ \bibinfo
  {author} {\bibfnamefont {K.~E.}\ \bibnamefont {Nagaev}},\ }\href {\doibase
  10.1103/PhysRevB.76.165426} {\bibfield  {journal} {\bibinfo  {journal} {Phys.
  Rev. B}\ }\textbf {\bibinfo {volume} {76}},\ \bibinfo {pages} {165426}
  (\bibinfo {year} {2007})}\BibitemShut {NoStop}%
\bibitem [{\citenamefont {Zgirski}\ \emph {et~al.}(2019)\citenamefont
  {Zgirski}, \citenamefont {Foltyn}, \citenamefont {Savin},\ and\ \citenamefont
  {Norowski}}]{Zgirski2019}%
  \BibitemOpen
  \bibfield  {author} {\bibinfo {author} {\bibfnamefont {M.}~\bibnamefont
  {Zgirski}}, \bibinfo {author} {\bibfnamefont {M.}~\bibnamefont {Foltyn}},
  \bibinfo {author} {\bibfnamefont {A.}~\bibnamefont {Savin}}, \ and\ \bibinfo
  {author} {\bibfnamefont {K.}~\bibnamefont {Norowski}},\ }\href {\doibase
  10.1103/PhysRevApplied.11.054070} {\bibfield  {journal} {\bibinfo  {journal}
  {Phys. Rev. Applied}\ }\textbf {\bibinfo {volume} {11}},\ \bibinfo {pages}
  {054070} (\bibinfo {year} {2019})}\BibitemShut {NoStop}%
\bibitem [{Sup()}]{Supplemental_Hot_electrons}%
  \BibitemOpen
  \href@noop {} {}\bibinfo {note} {See Supplemental Material for sample
  fabrication, experimental apparatus, metrological aspects of the switching
  thermometry, long heating pulse response, modeling of the heat flow,
  influence of the electron-phonon interaction on the quasiparticle diffusion,
  diffusion constant for various temperatures and different heating power,
  numerical calculations and material parameters.}\BibitemShut {Stop}%
\bibitem [{\citenamefont {Bezuglyi}\ and\ \citenamefont
  {Vinokur}(2003)}]{Vinokur2003}%
  \BibitemOpen
  \bibfield  {author} {\bibinfo {author} {\bibfnamefont {E.~V.}\ \bibnamefont
  {Bezuglyi}}\ and\ \bibinfo {author} {\bibfnamefont {V.}~\bibnamefont
  {Vinokur}},\ }\href {\doibase 10.1103/PhysRevLett.91.137002} {\bibfield
  {journal} {\bibinfo  {journal} {Phys. Rev. Lett.}\ }\textbf {\bibinfo
  {volume} {91}},\ \bibinfo {pages} {137002} (\bibinfo {year}
  {2003})}\BibitemShut {NoStop}%
\bibitem [{\citenamefont {Eom}\ \emph {et~al.}(1998)\citenamefont {Eom},
  \citenamefont {Chien},\ and\ \citenamefont
  {Chandrasekhar}}]{Chandrasekhar1998}%
  \BibitemOpen
  \bibfield  {author} {\bibinfo {author} {\bibfnamefont {J.}~\bibnamefont
  {Eom}}, \bibinfo {author} {\bibfnamefont {C.-J.}\ \bibnamefont {Chien}}, \
  and\ \bibinfo {author} {\bibfnamefont {V.}~\bibnamefont {Chandrasekhar}},\
  }\href {\doibase 10.1103/PhysRevLett.81.437} {\bibfield  {journal} {\bibinfo
  {journal} {Phys. Rev. Lett.}\ }\textbf {\bibinfo {volume} {81}},\ \bibinfo
  {pages} {437} (\bibinfo {year} {1998})}\BibitemShut {NoStop}%
\bibitem [{\citenamefont {Aumentado}\ \emph {et~al.}(2004)\citenamefont
  {Aumentado}, \citenamefont {Keller}, \citenamefont {Martinis},\ and\
  \citenamefont {Devoret}}]{Aumentado2004}%
  \BibitemOpen
  \bibfield  {author} {\bibinfo {author} {\bibfnamefont {J.}~\bibnamefont
  {Aumentado}}, \bibinfo {author} {\bibfnamefont {M.~W.}\ \bibnamefont
  {Keller}}, \bibinfo {author} {\bibfnamefont {J.~M.}\ \bibnamefont
  {Martinis}}, \ and\ \bibinfo {author} {\bibfnamefont {M.~H.}\ \bibnamefont
  {Devoret}},\ }\href {\doibase 10.1103/PhysRevLett.92.066802} {\bibfield
  {journal} {\bibinfo  {journal} {Phys. Rev. Lett.}\ }\textbf {\bibinfo
  {volume} {92}},\ \bibinfo {pages} {066802} (\bibinfo {year}
  {2004})}\BibitemShut {NoStop}%
\bibitem [{\citenamefont {Taupin}\ \emph {et~al.}(2016)\citenamefont {Taupin},
  \citenamefont {Khaymovich}, \citenamefont {Meschke}, \citenamefont
  {Mel'nikov},\ and\ \citenamefont {Pekola}}]{Taupin2016}%
  \BibitemOpen
  \bibfield  {author} {\bibinfo {author} {\bibfnamefont {M.}~\bibnamefont
  {Taupin}}, \bibinfo {author} {\bibfnamefont {I.~M.}\ \bibnamefont
  {Khaymovich}}, \bibinfo {author} {\bibfnamefont {M.}~\bibnamefont {Meschke}},
  \bibinfo {author} {\bibfnamefont {A.~S.}\ \bibnamefont {Mel'nikov}}, \ and\
  \bibinfo {author} {\bibfnamefont {J.~P.}\ \bibnamefont {Pekola}},\ }\href
  {\doibase 10.1038/ncomms10977} {\bibfield  {journal} {\bibinfo  {journal}
  {Nat. Commun.}\ }\textbf {\bibinfo {volume} {7}},\ \bibinfo {pages} {10977}
  (\bibinfo {year} {2016})}\BibitemShut {NoStop}%
\bibitem [{\citenamefont {Pekola}\ \emph {et~al.}(2000)\citenamefont {Pekola},
  \citenamefont {Anghel}, \citenamefont {Suppula}, \citenamefont {Suoknuuti},
  \citenamefont {Manninen},\ and\ \citenamefont {Manninen}}]{Manninen2000}%
  \BibitemOpen
  \bibfield  {author} {\bibinfo {author} {\bibfnamefont {J.~P.}\ \bibnamefont
  {Pekola}}, \bibinfo {author} {\bibfnamefont {D.~V.}\ \bibnamefont {Anghel}},
  \bibinfo {author} {\bibfnamefont {T.~I.}\ \bibnamefont {Suppula}}, \bibinfo
  {author} {\bibfnamefont {J.~K.}\ \bibnamefont {Suoknuuti}}, \bibinfo {author}
  {\bibfnamefont {A.~J.}\ \bibnamefont {Manninen}}, \ and\ \bibinfo {author}
  {\bibfnamefont {M.}~\bibnamefont {Manninen}},\ }\href {\doibase
  10.1063/1.126474} {\bibfield  {journal} {\bibinfo  {journal} {Applied Physics
  Letters}\ }\textbf {\bibinfo {volume} {76}},\ \bibinfo {pages} {2782}
  (\bibinfo {year} {2000})}\BibitemShut {NoStop}%
\bibitem [{\citenamefont {Martinis}\ \emph {et~al.}(2009)\citenamefont
  {Martinis}, \citenamefont {Ansmann},\ and\ \citenamefont
  {Aumentado}}]{Martinis2009}%
  \BibitemOpen
  \bibfield  {author} {\bibinfo {author} {\bibfnamefont {J.~M.}\ \bibnamefont
  {Martinis}}, \bibinfo {author} {\bibfnamefont {M.}~\bibnamefont {Ansmann}}, \
  and\ \bibinfo {author} {\bibfnamefont {J.}~\bibnamefont {Aumentado}},\ }\href
  {\doibase 10.1103/PhysRevLett.103.097002} {\bibfield  {journal} {\bibinfo
  {journal} {Phys. Rev. Lett.}\ }\textbf {\bibinfo {volume} {103}},\ \bibinfo
  {pages} {097002} (\bibinfo {year} {2009})}\BibitemShut {NoStop}%
\bibitem [{\citenamefont {Ullom}\ \emph {et~al.}(2000)\citenamefont {Ullom},
  \citenamefont {Fisher},\ and\ \citenamefont {Nahum}}]{Ullom2000}%
  \BibitemOpen
  \bibfield  {author} {\bibinfo {author} {\bibfnamefont {J.~N.}\ \bibnamefont
  {Ullom}}, \bibinfo {author} {\bibfnamefont {P.~A.}\ \bibnamefont {Fisher}}, \
  and\ \bibinfo {author} {\bibfnamefont {M.}~\bibnamefont {Nahum}},\ }\href
  {\doibase 10.1103/PhysRevB.61.14839} {\bibfield  {journal} {\bibinfo
  {journal} {Phys. Rev. B}\ }\textbf {\bibinfo {volume} {61}},\ \bibinfo
  {pages} {14839} (\bibinfo {year} {2000})}\BibitemShut {NoStop}%
\bibitem [{\citenamefont {Riwar}\ \emph {et~al.}(2016)\citenamefont {Riwar},
  \citenamefont {Hosseinkhani}, \citenamefont {Burkhart}, \citenamefont {Gao},
  \citenamefont {Schoelkopf}, \citenamefont {Glazman},\ and\ \citenamefont
  {Catelani}}]{Riwar2016}%
  \BibitemOpen
  \bibfield  {author} {\bibinfo {author} {\bibfnamefont {R.-P.}\ \bibnamefont
  {Riwar}}, \bibinfo {author} {\bibfnamefont {A.}~\bibnamefont {Hosseinkhani}},
  \bibinfo {author} {\bibfnamefont {L.~D.}\ \bibnamefont {Burkhart}}, \bibinfo
  {author} {\bibfnamefont {Y.~Y.}\ \bibnamefont {Gao}}, \bibinfo {author}
  {\bibfnamefont {R.~J.}\ \bibnamefont {Schoelkopf}}, \bibinfo {author}
  {\bibfnamefont {L.~I.}\ \bibnamefont {Glazman}}, \ and\ \bibinfo {author}
  {\bibfnamefont {G.}~\bibnamefont {Catelani}},\ }\href {\doibase
  10.1103/PhysRevB.94.104516} {\bibfield  {journal} {\bibinfo  {journal} {Phys.
  Rev. B}\ }\textbf {\bibinfo {volume} {94}},\ \bibinfo {pages} {104516}
  (\bibinfo {year} {2016})}\BibitemShut {NoStop}%
\bibitem [{\citenamefont {Rajauria}\ \emph {et~al.}(2012)\citenamefont
  {Rajauria}, \citenamefont {Pascal}, \citenamefont {Gandit}, \citenamefont
  {Hekking}, \citenamefont {Pannetier},\ and\ \citenamefont
  {Courtois}}]{Courtois2012}%
  \BibitemOpen
  \bibfield  {author} {\bibinfo {author} {\bibfnamefont {S.}~\bibnamefont
  {Rajauria}}, \bibinfo {author} {\bibfnamefont {L.~M.~A.}\ \bibnamefont
  {Pascal}}, \bibinfo {author} {\bibfnamefont {P.}~\bibnamefont {Gandit}},
  \bibinfo {author} {\bibfnamefont {F.~W.~J.}\ \bibnamefont {Hekking}},
  \bibinfo {author} {\bibfnamefont {B.}~\bibnamefont {Pannetier}}, \ and\
  \bibinfo {author} {\bibfnamefont {H.}~\bibnamefont {Courtois}},\ }\href
  {\doibase 10.1103/PhysRevB.85.020505} {\bibfield  {journal} {\bibinfo
  {journal} {Phys. Rev. B}\ }\textbf {\bibinfo {volume} {85}},\ \bibinfo
  {pages} {020505} (\bibinfo {year} {2012})}\BibitemShut {NoStop}%
\bibitem [{\citenamefont {Stan}\ \emph {et~al.}(2004)\citenamefont {Stan},
  \citenamefont {Field},\ and\ \citenamefont {Martinis}}]{Martinis2004}%
  \BibitemOpen
  \bibfield  {author} {\bibinfo {author} {\bibfnamefont {G.}~\bibnamefont
  {Stan}}, \bibinfo {author} {\bibfnamefont {S.~B.}\ \bibnamefont {Field}}, \
  and\ \bibinfo {author} {\bibfnamefont {J.~M.}\ \bibnamefont {Martinis}},\
  }\href {\doibase 10.1103/PhysRevLett.92.097003} {\bibfield  {journal}
  {\bibinfo  {journal} {Phys. Rev. Lett.}\ }\textbf {\bibinfo {volume} {92}},\
  \bibinfo {pages} {097003} (\bibinfo {year} {2004})}\BibitemShut {NoStop}%
\end{thebibliography}

\begin{thebibliography}{5}%
\makeatletter
\providecommand \@ifxundefined [1]{%
 \@ifx{#1\undefined}
}%
\providecommand \@ifnum [1]{%
 \ifnum #1\expandafter \@firstoftwo
 \else \expandafter \@secondoftwo
 \fi
}%
\providecommand \@ifx [1]{%
 \ifx #1\expandafter \@firstoftwo
 \else \expandafter \@secondoftwo
 \fi
}%
\providecommand \natexlab [1]{#1}%
\providecommand \enquote  [1]{``#1''}%
\providecommand \bibnamefont  [1]{#1}%
\providecommand \bibfnamefont [1]{#1}%
\providecommand \citenamefont [1]{#1}%
\providecommand \href@noop [0]{\@secondoftwo}%
\providecommand \href [0]{\begingroup \@sanitize@url \@href}%
\providecommand \@href[1]{\@@startlink{#1}\@@href}%
\providecommand \@@href[1]{\endgroup#1\@@endlink}%
\providecommand \@sanitize@url [0]{\catcode `\\12\catcode `\$12\catcode
  `\&12\catcode `\#12\catcode `\^12\catcode `\_12\catcode `\%12\relax}%
\providecommand \@@startlink[1]{}%
\providecommand \@@endlink[0]{}%
\providecommand \url  [0]{\begingroup\@sanitize@url \@url }%
\providecommand \@url [1]{\endgroup\@href {#1}{\urlprefix }}%
\providecommand \urlprefix  [0]{URL }%
\providecommand \Eprint [0]{\href }%
\providecommand \doibase [0]{http://dx.doi.org/}%
\providecommand \selectlanguage [0]{\@gobble}%
\providecommand \bibinfo  [0]{\@secondoftwo}%
\providecommand \bibfield  [0]{\@secondoftwo}%
\providecommand \translation [1]{[#1]}%
\providecommand \BibitemOpen [0]{}%
\providecommand \bibitemStop [0]{}%
\providecommand \bibitemNoStop [0]{.\EOS\space}%
\providecommand \EOS [0]{\spacefactor3000\relax}%
\providecommand \BibitemShut  [1]{\csname bibitem#1\endcsname}%
\let\auto@bib@innerbib\@empty
\bibitem [{\citenamefont {Zgirski}\ \emph {et~al.}(2018)\citenamefont
  {Zgirski}, \citenamefont {Foltyn}, \citenamefont {Savin}, \citenamefont
  {Norowski}, \citenamefont {Meschke},\ and\ \citenamefont
  {Pekola}}]{Zgirski2018_S}%
  \BibitemOpen
  \bibfield  {author} {\bibinfo {author} {\bibfnamefont {M.}~\bibnamefont
  {Zgirski}}, \bibinfo {author} {\bibfnamefont {M.}~\bibnamefont {Foltyn}},
  \bibinfo {author} {\bibfnamefont {A.}~\bibnamefont {Savin}}, \bibinfo
  {author} {\bibfnamefont {K.}~\bibnamefont {Norowski}}, \bibinfo {author}
  {\bibfnamefont {M.}~\bibnamefont {Meschke}}, \ and\ \bibinfo {author}
  {\bibfnamefont {J.}~\bibnamefont {Pekola}},\ }\href {\doibase
  10.1103/PhysRevApplied.10.044068} {\bibfield  {journal} {\bibinfo  {journal}
  {Phys. Rev. Applied}\ }\textbf {\bibinfo {volume} {10}},\ \bibinfo {pages}
  {044068} (\bibinfo {year} {2018})}\BibitemShut {NoStop}%
\bibitem [{\citenamefont {Zgirski}\ \emph {et~al.}(2019)\citenamefont
  {Zgirski}, \citenamefont {Foltyn}, \citenamefont {Savin},\ and\ \citenamefont
  {Norowski}}]{Zgirski2019_S}%
  \BibitemOpen
  \bibfield  {author} {\bibinfo {author} {\bibfnamefont {M.}~\bibnamefont
  {Zgirski}}, \bibinfo {author} {\bibfnamefont {M.}~\bibnamefont {Foltyn}},
  \bibinfo {author} {\bibfnamefont {A.}~\bibnamefont {Savin}}, \ and\ \bibinfo
  {author} {\bibfnamefont {K.}~\bibnamefont {Norowski}},\ }\href {\doibase
  10.1103/PhysRevApplied.11.054070} {\bibfield  {journal} {\bibinfo  {journal}
  {Phys. Rev. Applied}\ }\textbf {\bibinfo {volume} {11}},\ \bibinfo {pages}
  {054070} (\bibinfo {year} {2019})}\BibitemShut {NoStop}%
\bibitem [{\citenamefont {Maisi}\ \emph {et~al.}(2013)\citenamefont {Maisi},
  \citenamefont {Lotkhov}, \citenamefont {Kemppinen}, \citenamefont {Heimes},
  \citenamefont {Muhonen},\ and\ \citenamefont {Pekola}}]{Maisi2013_S}%
  \BibitemOpen
  \bibfield  {author} {\bibinfo {author} {\bibfnamefont {V.~F.}\ \bibnamefont
  {Maisi}}, \bibinfo {author} {\bibfnamefont {S.~V.}\ \bibnamefont {Lotkhov}},
  \bibinfo {author} {\bibfnamefont {A.}~\bibnamefont {Kemppinen}}, \bibinfo
  {author} {\bibfnamefont {A.}~\bibnamefont {Heimes}}, \bibinfo {author}
  {\bibfnamefont {J.~T.}\ \bibnamefont {Muhonen}}, \ and\ \bibinfo {author}
  {\bibfnamefont {J.~P.}\ \bibnamefont {Pekola}},\ }\href {\doibase
  10.1103/PhysRevLett.111.147001} {\bibfield  {journal} {\bibinfo  {journal}
  {Phys. Rev. Lett.}\ }\textbf {\bibinfo {volume} {111}},\ \bibinfo {pages}
  {147001} (\bibinfo {year} {2013})}\BibitemShut {NoStop}%
\bibitem [{\citenamefont {Courtois}\ \emph {et~al.}(2008)\citenamefont
  {Courtois}, \citenamefont {Meschke}, \citenamefont {Peltonen},\ and\
  \citenamefont {Pekola}}]{Courtois2008_S}%
  \BibitemOpen
  \bibfield  {author} {\bibinfo {author} {\bibfnamefont {H.}~\bibnamefont
  {Courtois}}, \bibinfo {author} {\bibfnamefont {M.}~\bibnamefont {Meschke}},
  \bibinfo {author} {\bibfnamefont {J.~T.}\ \bibnamefont {Peltonen}}, \ and\
  \bibinfo {author} {\bibfnamefont {J.~P.}\ \bibnamefont {Pekola}},\ }\href
  {\doibase 10.1103/PhysRevLett.101.067002} {\bibfield  {journal} {\bibinfo
  {journal} {Phys. Rev. Lett.}\ }\textbf {\bibinfo {volume} {101}},\ \bibinfo
  {pages} {067002} (\bibinfo {year} {2008})}\BibitemShut {NoStop}%
\bibitem [{\citenamefont {Phillips}(1959)}]{Philips1959_S}%
  \BibitemOpen
  \bibfield  {author} {\bibinfo {author} {\bibfnamefont {N.~E.}\ \bibnamefont
  {Phillips}},\ }\href {\doibase 10.1103/PhysRev.114.676} {\bibfield  {journal}
  {\bibinfo  {journal} {Phys. Rev.}\ }\textbf {\bibinfo {volume} {114}},\
  \bibinfo {pages} {676} (\bibinfo {year} {1959})}\BibitemShut {NoStop}%
\end{thebibliography}
\end{document}